\documentclass[aps,prc,twocolumn,superscriptaddress,showpacs,floatfix]{revtex4-1}

\usepackage[T1]{fontenc}
\usepackage{latexsym}
\usepackage{graphicx,amssymb}
\usepackage[fleqn]{amsmath}
\usepackage{amsfonts}
\usepackage{dcolumn}% Align table columns on decimal point
\usepackage{bm}% bold math
\usepackage{braket}
\usepackage{multirow}
\usepackage{MnSymbol}
\usepackage{color}
\usepackage{ulem}
\usepackage{hyperref}
\usepackage{bm}% bold math

\begin{document}

\preprint{ }

\title{$^{14}$N(p,$\gamma)^{15}$O $S$ factor and the puzzling solar composition problem}

\author{G.X. Dong}
\affiliation{School of Science, Huzhou University, Huzhou 313000, China}

\author{X.B. Wang}
\affiliation{School of Science, Huzhou University, Huzhou 313000, China}

\author{N. Michel}
\affiliation{CAS Key Laboratory of High Precision Nuclear Spectroscopy,Institute of Modern Physics, Chinese Academy of Sciences, Lanzhou 730000, China}
\affiliation{School of Nuclear Science and Technology, University of Chinese Academy of Sciences, Beijing 100049, China}

\author{M. P{\l}oszajczak}
\thanks{marek.ploszajczak@ganil.fr}
\affiliation{Grand Acc\'el\'erateur National d'Ions Lourds (GANIL), CEA/DSM - CNRS/IN2P3,
BP 55027, F-14076 Caen Cedex, France}

\date{\today}

\begin{abstract}
In stellar hydrogen burning, the CNO cycle dominates, with the $^{14}$N(p,$\gamma)^{15}$O  reaction being the slowest process. Consequently, this reaction critically influences the solar composition, CNO neutrino fluxes, and the evolution of star clusters and galaxies. Recent direct measurements of $^{14}$N(p,$\gamma)^{15}$O  have reported an enhanced astrophysical $S$-factor. This work presents a microscopic theoretical study of the $^{14}$N(p,$\gamma)^{15}$O  reaction using the Gamow shell model in the coupled-channel representation (GSM-CC). The calculations achieve good agreement with experimental data for both the total $S$-factors and the separate contributions from transitions to the ground state and excited states of $^{15}\mathrm{O}$. However, the predicted $S$-factor at zero energy exceeds the experimental value. Based on the computed $S$-factors, the derived carbon and nitrogen abundances align closely with predictions from recent $^{14}$N(p,$\gamma)^{15}$O  cross-section measurements, yet remain significantly lower than the latest solar neutrino observation values.
\end{abstract}

\maketitle

\textit{Introduction --} The key input of the standard solar model is the solar metallicity, which is evaluated by the ratio of elements heavier than helium to hydrogen in the Sun~\cite{Bahcall04,Bahcall_2006}. The early estimation of solar metallicity was derived from the spectra of the solar photosphere, which agreed with helioseimology measurements~\cite{Bahcall_2005}. Later, using more advanced techniques~\footnote{The main improvement of the technique stems from improved atomic and molecular transition probabilities, the development of 3D hydrodynamical solar model atmosphere, and non-local thermodynamic equilibrium (LTE) in the spectral line formation~\cite{Asplund09,Caffau2011,Asplund2021}.}, a 30\%--40\% reduction~\cite{Asplund09,Caffau2011,Asplund2021} of solar metallicity has been found as compared to the traditional modeling~\cite{Grevesse1998}.
This contradiction, the so-called ``solar composition problem'', is the long-standing puzzle for over twenty years.

In the Sun, $\sim$2\% of the energy is produced by the carbon-nitrogen-oxygen (CNO) cycle~\cite{Haxton_2008}. For stars about 1.3 times more massive than the Sun, this process dominates energy production~\cite{Wiescher18}, and its importance depends on metallicity. $^{14}$N(p,$\gamma)^{15}$O is the slowest process in the CNO cycle. Therefore, a precise determination of the $^{14}$N(p,$\gamma)^{15}$O astrophysical $S$ factor is important not only for studying the composition of the Sun and CNO neutrinos, but also for understanding the evolution and age of star clusters and galaxies.
%and thus its reaction rate has large impact on the CN abundance and energy production of CNO cycle.
Solar neutrinos provide direct information about the interior of the Sun, making them an effective tool for solving the problem of the Sun's composition.
Recently, the solar neutrinos from the CNO cycle were measured by Borexino collaboration~\cite{14NNeutrinos2020,14NNeutrinos2022}
providing important insight about the solar metallicity. The observed flux of CNO neutrinos from the decay of $^{15}$O, which is directly related with the rate of the reaction $^{14}$N(p,$\gamma)^{15}$O and, hence, the astrophysical $S$ factor, suggests the enhanced production rate of $^{15}$O.

A lot of efforts have been made to measure the  $^{14}$N(p,$\gamma)^{15}$O reaction cross section. Despite that, still the significant differences remain between various experiments concerning both the total zero energy astrophysical factor S$(0)$ and the individual contributions to the S$(0)$ of the transitions to the ground state and excited states of $^{15}$O.
%In 1987, the first comprehensive measurements were made by Schr\"oder et. al.~\cite{SCHRODER1987240}, for the proton energy from 0.2 to 3.6 MeV. The S factor at zero energy S$(0)$ was expected to be very large, as 3.20$\pm$0.54 keV barn. However, it was later realized that their analysis of data had errors and should be corrected~\cite{ANGULO2001755}, thus a factor of 1.7 reduction was deduced.
Several measurements, extending to lower energies, have been performed at LUNA underground laboratory~\cite{FORMICOLA200461,Imbriani200514N,Bemmerer200614N,LEMUT2006483,Marta0814N}.
Direct measurement at LENA facility~\cite{Runkle200514N} found an almost twice as large astrophysical factor at zero energy for the transition to the ground state S$_{g.s.}(0)$ than previously obtained at LUNA. Five angles of differential cross sections were measured by Li \textit{et al.} at Notre Dame for $E_p$ = 0.7 to 3.6 MeV~\cite{Li201614N}. Other experiments done at the CASPAR underground accelerator~\cite{Frentz202214N} and the HZDR tandem accelerator~\cite{Wagner201814N}, provided additional constraints on the $R$-matrix fitting. The total non-resonant S factors were also measured with the activation method in the range of center-of-mass (c.m.) energies from 0.55 to 1.4 MeV~\cite{ATOMKI14N}.
Last year, the new Solar Fusion III Compilation (SF-III) has recommended the value S$(0)$=1.68 $\pm$ 0.14 keV b~\cite{SF-III}
However, the recent measurement at HINEG facility in Hefei
%Hefei 350 kV accelerator facility
of the S-factors for all transitions in the energy range $E_p = 0.11 - 0.26$ MeV~\cite{chen202414npg} gave a value of S$(0)$=1.92$\pm$0.08 keV b, which is 14\% higher than the recommended value. This value of S$(0)$ implies a high metallicity composition for the Sun, although it is also consistent with a low metallicity composition with a confidence level of 1$\sigma$.

At energies corresponding to solar temperatures, direct measurement of the $^{14}$N(p,$\gamma)^{15}$O reaction cross section is impossible under present experimental conditions. In this case, theoretical studies are necessary. Theoretical studies of $^{14}$N(p,$\gamma)^{15}$O reaction were done mainly using the $R$-matrix method~\cite{ANGULO2001755,Mukhamedzhanov2003,Imbriani200514N,Bemmerer200614N,Runkle200514N,AZURE,Li201614N,Wagner201814N,Frentz2021lifetime,Frentz202214N,chen202414npg}, or the modified potential cluster model~\cite{Dubovichenko14N}.
In this work, we present the first microscopic study of this reaction using the Gamow shell model in the coupled-channels representation (GSM-CC)~\cite{GSMbook}. The GSM-CC provides a unified theoretical framework for nuclear structure and reaction studies. It has been applied with the success to describe nuclear spectra, proton elastic and inelastic scattering~\cite{Jaganathen14}, deuteron elastic scattering~\cite{Mercenne19}, triton and $^3$He elastic scattering~\cite{Fernandez23}, and various radiative capture reactions~\cite{Fossez15,dong17,dong22,dong23,DongGX2024}.

\vskip 0.2 truecm
\textit{Gamow shell model in the coupled-channel representation  --} Here, we briefly introduce the essential steps of the GSM-CC approach. More details can be found in the Supplemental Material~\cite{supplement} (including references~\cite{GSMbook,Michel09,Ikeda88,rf:4,Michel02RIB,Michel03,Mercenne19,Furutani78,Furutani79,jaganathen_2017,XuHM1994,Burjan20,Mukhamedzhanov2003,FORMICOLA200461,Bertone2002,Li201614N,ANGULO2001755}).

In GSM-CC, the ${ A }$-body system is described by the reaction channels:%
\begin{equation}
  \ket{ { \Psi }_{ M }^{ J } } = \sum_{ {c} } \int_{ 0 }^{ +\infty } \ket{{ \left( r,c \right) }_{ M }^{ J } } \frac{ { u }_{ {c} }^{JM} (r) }{ r } { r }^{ 2 } ~ dr \; ,
  \label{scat_A_body_compound}
\end{equation}
%
%\textcolor{red}
{
where ${ {u}_{ {c} }^{JM}(r) }$ is the radial amplitude of channel $c$, and ${ r }$ is the cluster orbital shell model (COSM) radial coordinate of the projectile,
defined with respect to the inert $^4$He core (see Supplement~\cite{supplement} for details about the COSM formalism).
}
${ {u}_{ {c} }^{JM}(r) }$ is obtained by solving the coupled-channel equations for fixed total angular momentum $ {J} $ and projection $ {M} $.
$\ket{ \left( r,c \right)^J_M} $ denotes the binary-cluster channel:
\begin{equation}
  \ket{ \left( r,c \right)^J_M}  = \hat{ \mathcal{A}} \ket{ \{ \ket{ \Psi_{ {\rm T} }^{J_{ {\rm T} } }  }
  \otimes \ket{ {\Psi^{J_{{\rm P}}}_{{\rm P}}}} \}_{ M }^{J} } \; .
  \label{channel}
\end{equation}
The channel index $c$ contains information about binary mass partition and quantum numbers, and the variable $r$ is included in $\ket{\Psi_{\rm P}^{J_{\rm P}}}$. ${\hat{ \mathcal{A}}}$ stands for the antisymmetrization of nucleons in different clusters. $\ket{\Psi_{\rm T}^{J_{\rm T}}}$ and $\ket{\Psi_{\rm P}^{J_{\rm P}} }$ denote the target and projectile states with their respective angular momenta ${ { J }_{ {\rm T} } }$ and ${ { J }_{ {\rm P} } }$. The total angular momentum ${  J }$ is obtained by coupling target and projectile angular momenta ${ { J }_{ {\rm T} } }$ and ${ { J }_{ {\rm P} } }$.

In the calculation of the cross section of the $^{14}$N(p,$\gamma)^{15}$O reaction, the one-body part of the Hamiltonian is represented by a Woods-Saxon potential. For the two-body interaction, the Furutani-Horiuchi-Tamagaki finite-range two-body force is used~\cite{Furutani78,Furutani79}. More details of the one-body potential and the two-body interaction between valence particles are given in the Supplement~\cite{supplement}. The model spaces for the target $^{14}$N and the reaction composite $^{15}$O are presented in the following two paragraphs.

For the $^{14}$N target wave functions in GSM, the model space for the protons consists of $psd$ partial waves. The $p_{1/2}$ and $p_{3/2}$ partial waves are represented by six harmonic oscillator (HO) states to discretize each $\ell_j$ scattering continuum. For HO states, the oscillator length was assumed to be 2.0 fm. For $sd$ partial waves, $0d_{5/2}$, $0d_{3/2}$,  $1s_{1/2}$ resonant single-particle (s.p.) state, and 30 non-resonant s.p.~continuum states in each Berggren contour $\mathcal{L}^+_{d_{5/2}}$, $\mathcal{L}^+_{d_{3/2}}$,  and $\mathcal{L}^+_{s_{1/2}}$ are used. Each contour contains three segments connecting the points: $k_{\text{min}}$=0.0, $k_{\text{peak}}=0.15-i0.10$ fm$^{-1}$, $k_{\text{mid}}=0.3$ fm$^{-1}$ and $k_{\text{max}}$=2.0 fm$^{-1}$. For neutrons,  $0p_{3/2}$, $0p_{1/2}$, $0d_{5/2}$, $0d_{3/2}$, and $1s_{1/2}$ s.p.~states are included. The $0s_{1/2}$ shell which is fully occupied does not belong to the model space. One proton or one neutron in the $0p_{3/2}$ shell can be excited to other shells. Thus, the GSM calculations are performed in 105 shells for the proton, and 5 shells for the neutron.

The GSM-CC channel states in $^{15}$O are then built by coupling the low-lying GSM states of $^{14}$N with the proton in partial waves: $p_{3/2}$, $p_{1/2}$, $d_{5/2}$, $s_{1/2}$ and $d_{3/2}$. When constructing the reaction channels, the following states with positive parity 0$^+_1, $1$^+_1$, 1$^+_2$ and negative parity 0$^-_1$, 1$^-_1$, 1$^-_2$, 1$^-_3$, 2$^-_1$, 3$^-_1$ in $^{14}$N are taken into account. For proton partial waves, s.p.~resonant states $0p_{3/2}$, $0p_{1/2}$, $0d_{5/2}$, $0d_{3/2}$, $1s_{1/2}$ and 60 non-resonant s.p.~continuum states in each Berggren contour $\mathcal{L}^+_{p_{1/2}}$, $\mathcal{L}^+_{p_{3/2}}$, $\mathcal{L}^+_{d_{5/2}}$, $\mathcal{L}^+_{d_{3/2}}$ and $\mathcal{L}^+_{s_{1/2}}$ are considered.
Three segments connecting points $k_{\text{min}}$=0.0, $k_{\text{peak}}=0.15-i0.01$ fm$^{-1}$, $k_{\text{mid}}=0.3-i0.01$ fm$^{-1}$ and $k_{\text{max}}$=2.0 fm$^{-1}$ are chosen for each contour. The total number of channel states for the $1/2^-$, $1/2^+$, $3/2^+$, $3/2^-$, $5/2^+$, and $7/2^+$ states in $^{15}$O are 15, 13, 17, 20, 12, and 5 respectively.

%\textcolor{red}
%{
 %Model space truncations must be performed because of the large number of valence nucleons. Indeed, one has 10 or 11 valence particles above the $^4$He inert core, so that the full model space is prohibitively large.
 %As the $0p_{3/2}$ and $0p_{1/2}$ shells are filled at Hartree-Fock level except for one state, it is sufficient to impose 1p-1h truncations from the $0p_{3/2}$ shell to the rest of the valence space.
 %Indeed, the 2p-2h, 3p-3h, etc, excitations from the $0p_{3/2}$ shells imply that $sd$ resonant or scattering shells must be occupied, so that they are energetically costly and can be renormalized in the Hamiltonian parameters.
 %Otherwise, up to 2p-2h excitations from one-body occupied resonant shells to the scattering continuum are otherwise demanded to reduce model space dimensions.
%}

\begin{figure}[htb]
\includegraphics[width=0.9\linewidth]{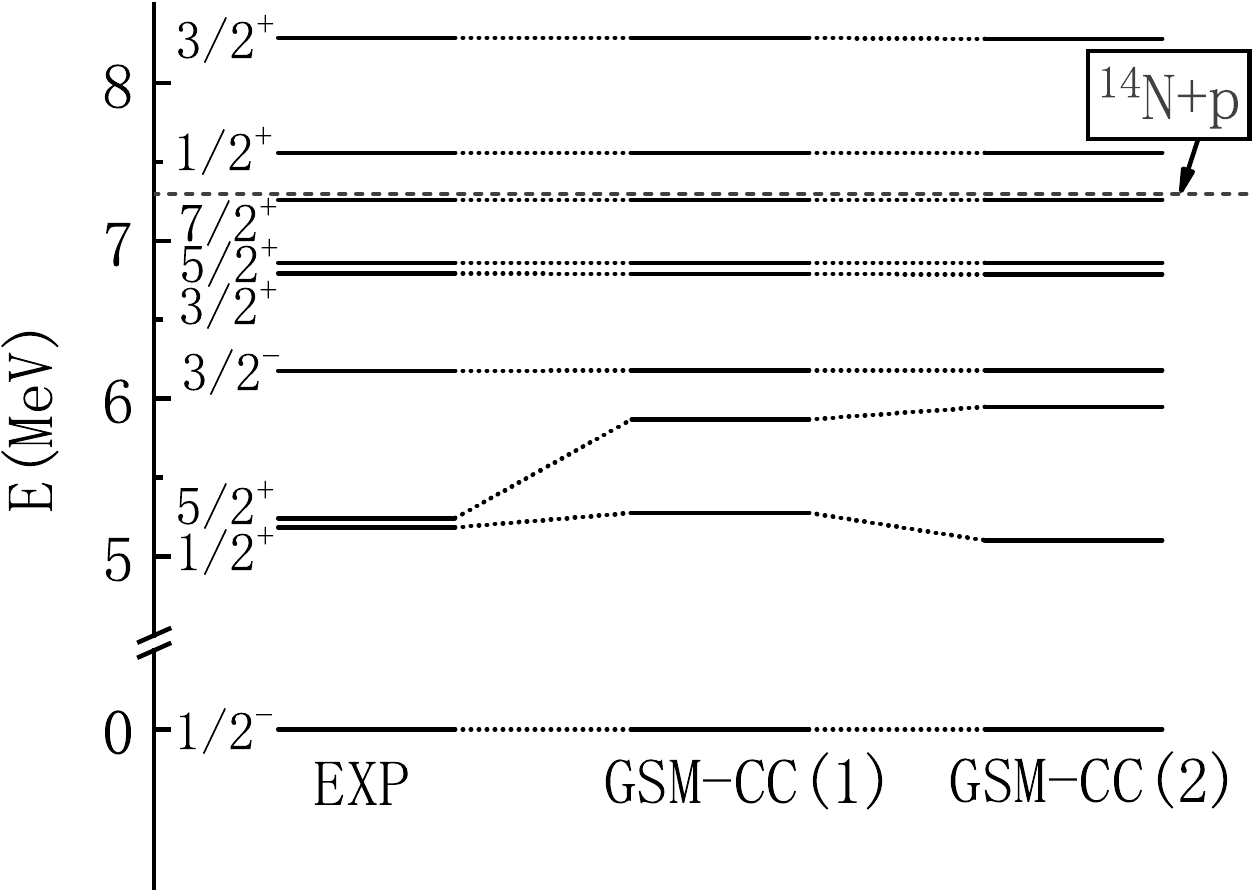}
\caption{The calculated energy levels of $^{15}$O are compared with the experimental data~\cite{nndc}. The ground state energy is set to zero and energies of the excited states are given relative to the ground state energy. The energy of the proton capture threshold is indicated by the dashed line. Two sets of GSM-CC calculations are shown, the details of which are explained in the text.}
\label{fig-1b}
\end{figure}

\vskip 0.2 truecm
\textit{Results --} GSM-CC and experimental energy levels of $^{15}$O which are important for the $^{14}$N(p,$\gamma)^{15}$O reaction cross section are shown in Fig.~\ref{fig-1b}. Two Hamiltonians are considered in the GSM-CC calculation which provide very similar energy spectra of $^{15}$O, with a notable exception for the $1/2^+_1$ and $5/2^+_1$ states.
In the GSM-CC(2) Hamiltonian, the depth of the Woods-Saxon potential for the $sd$ shells is shifted downwards by 2 MeV with respect to its value in the GSM-CC(1) Hamiltonian.
Moreover, the corrective factors in GSM-CC(2) are slightly readjusted with respect to GSM-CC(1) (see the Supplemental Material).

\begin{table}[htbp]
\caption{The calculated energies of the $^{13}$C ground state and $^{14}$N low-lying states. The unit is in MeV. The energies are relative to the ground state of $^{12}$C.}
\label{tab-energy}
\begin{ruledtabular}
\begin{tabular}{rrrr}
  & EXP~\cite{nndc} & GSM(1) & GSM(2) \\
\hline
$^{13}$C~~$1/2^{-}_1$ & -4.946 & -4.894 & -4.894 \\
\hline
$^{14}$N~~~~~$1^{+}_1$ & -12.497 & -12.548 & -12.595 \\
$0^{+}_1$ & -10.185 & -10.140 & -10.150 \\
$1^{+}_2$ & -8.549 & -7.640 & -7.657 \\
$0^{-}_1$ & -7.582 & -7.784 & -8.598 \\
\end{tabular}
\end{ruledtabular}
\end{table}

The agreement between theory and experiment is good, except for the state $5/2^+_1$. In GSM-CC(1), the calculated proton decay width $\Gamma(1/2^+_2) = 0.14$ keV of the $1/2^+_2$ resonance  is less than the experimental value of 0.99 keV~\cite{nndc}. For the $3/2^+_2$ resonance, the calculated width $\Gamma(3/2^+_2) = 6.6$ keV agrees reasonably well with the experimental value of 3.6 keV~\cite{nndc}. The calculated energies for the $^{13}$C ground state and the low-lying states of $^{14}$N are presented in Table~\ref{tab-energy}, showing good agreement with the experimental data. The binding energy of $^{14}$N relative to the $^{12}$C is -12.548 MeV, which is close to the experimental value of -12.497 MeV. In GSM-CC(2) calculations, the width of the $1/2^+_2$ resonance decreases to 0.07 keV, while the $3/2^+_2$ width becomes 5.89 keV. The $^{14}$N binding energy in GSM-CC(2) is -12.595 MeV. In the radiative capture cross section calculations, we use the exact $^{14}$N binding energy in the coupled-channel equations to properly account for the proton separation energy.
\begin{table}[ht!]
\caption{ The calculated magnetic  moments of low-lying states in $^{14}$N and $^{15}$O are compared with the experimental data taken from the compilation~\cite{stone2014table}. }
\label{m-moment}
\begin{ruledtabular}
\begin{tabular}{lccc}
  & Experiment~\cite{stone2014table} &GSM (1) &GSM (2)\\
    \hline
 $\mu \left(\mu_\textsc{N} \right)$  &  & & \\
  $^{14}$N, 1$^+_1$ & 0.40376100(6)  &0.317&0.317  \\
  $^{14}$N, 2$^-_1$  &1.32(8)   &1.086 &1.651 \\
  $^{14}$N, 3$^-_1$  & 2.0(5)  &1.721&2.080  \\
  $^{15}$O, 1/2$^-_1$  &0.71951(12)  &0.671 &0.679\\
\end{tabular}
\end{ruledtabular}
\end{table}
Additional information about the GSM-CC wave functions is given in Table ~\ref{m-moment}, which compares the calculated and experimental magnetic moments for the low-lying $^{14}$N and $^{15}$O states. One can see that the calculation reproduces the experimental data quite well.

\begin{figure}[htbp]
	\includegraphics[width=1.0\linewidth]{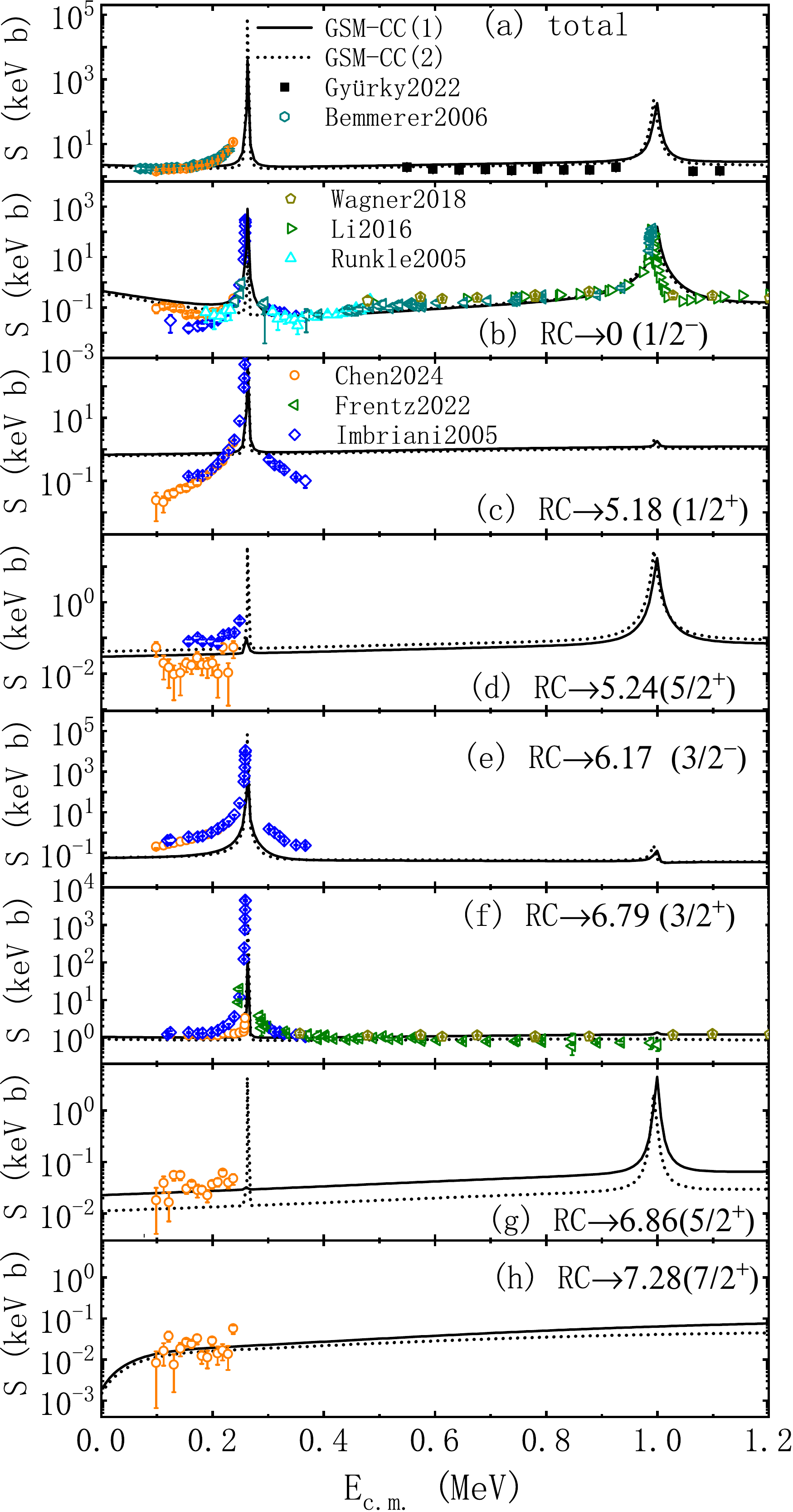}
	\caption{The astrophysical factor of the $^{14}$N(p,$\gamma)^{15}$O reaction is shown as the function of the proton projectile energy in the $p$ + $^{14}$N center of mass frame. The total S factor is given in Panel (a). The contributions by the capture to the ground state and to other excited states of $^{15}$O are plotted in Panel (b)-(h) respectively. The experimental data are listed for comparisons~\cite{Imbriani200514N,Runkle200514N,Bemmerer200614N,ATOMKI14N,Li201614N,Wagner201814N,Frentz202214N,chen202414npg}.  }
	\label{fig-2}
\end{figure}

Figure ~\ref{fig-2} shows the total astrophysical S-factor as well as the partial S-factors for radiative proton capture to both the ground state and the excited states of $^{15}$O. The figure shows results of the GSM-CC(1) and GSM-CC(2) calculations along with data from multiple experiments.
The calculated $S$ factors contain contributions from $E1$, $M1$ and $E2$ transitions. For $E1$ and $E2$ transitions, the effective charges obtained from the center of mass recoil correction~\cite{Hornyak75,YKHo88} were used.
The details of calculation of the antisymmetrized matrix element of the electromagnetic operators can be found in Ref.~\cite{Fossez15}. The infinite-range of the electromagnetic operator is fully treated by radial integrals of the GSM-CC matrix elements~\cite{Fossez15,dong17,GSMbook}. The short-range operators are calculated in the harmonic oscillator basis, as they are important only in the nuclear zone. Since the long-range operators extend in the asymptotic region, they are handled by the complex rotations in the radial integrals~\cite{Fossez15,dong17,GSMbook}.

\begin{table}[ht!]
\caption{The partial S factors at zero energy.}
\label{tab:S}
\begin{ruledtabular}
\begin{tabular}{ccc}
$J^{\pi}$ & GSM-CC(1) & GSM-CC(2) \\
\hline
$1/2^-_1$ & 0.447 & 0.365 \\
$1/2^+_1$ & 0.673 & 0.622 \\
$5/2^+_1$ & 0.029 & 0.041 \\
$3/2^-_1$ & 0.056 & 0.058 \\
$3/2^+_1$ & 1.033 & 0.897 \\
$5/2^+_2$ & 0.023 & 0.011 \\
$7/2^+_1$ & 0.002 & 0.001 \\
\end{tabular}
\end{ruledtabular}
\end{table}
The two GSM-CC calculations exhibit similar S factors, demonstrating good overall agreement with experimental data.
%Especially, the calculated total S factor in the low-energy range $E_{\mathrm{c.m.}} \approx 0.07 $--$ 0.23$ MeV is close to the experimental measurements~\cite{Bemmerer200614N,chen202414npg}.
The dominating contribution to the S factor is from the capture to the $3/2^+_1$ excited state at 6.79 MeV, and the GSM-CC results agree well with the experimental data. Significant discrepancies between different measurements of the astrophysical factor for the transition to the ground state~\cite{Runkle200514N,Marta0814N,Imbriani200514N,Frentz202214N,chen202414npg} occur in the very low energy range. The GSM-CC results for this transition show better agreement with the recent data~\cite{chen202414npg}.

\begin{figure}[htbp]
	\includegraphics[width=1.0\linewidth]{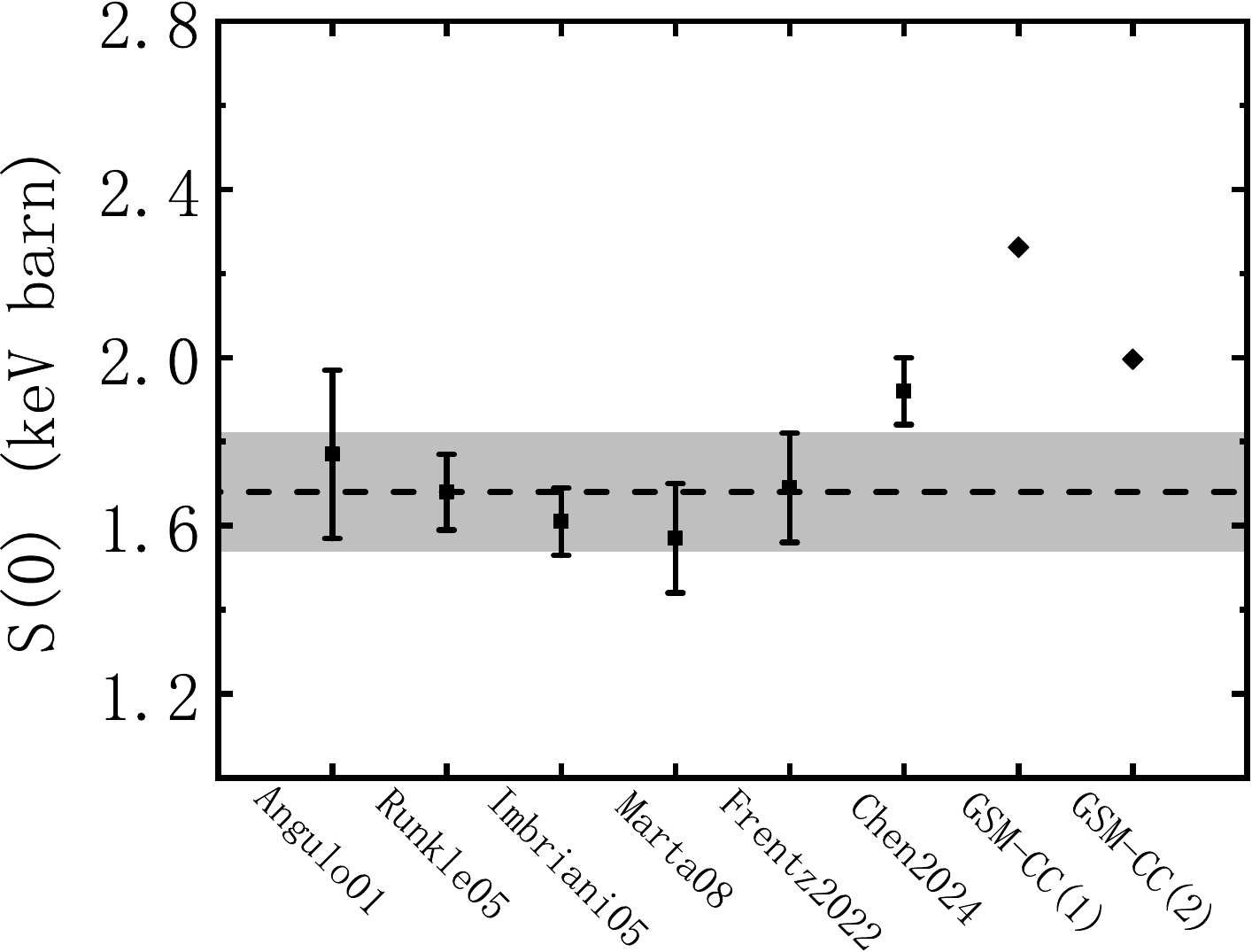}
	\caption{The comparison of S factors at the zero energy. The experimental data are taken from Refs.~\cite{ANGULO2001755,Runkle200514N,Marta0814N,Imbriani200514N,Frentz202214N,chen202414npg}. The shaded area is for the recommendation value in SF-III~\cite{SF-III}. }
	\label{fig-3}
\end{figure}

The GSM-CC energy of the $5/2^+_1$ state exhibits significant discrepancies with the experimental energy, as shown in Fig.~\ref{fig-1b}. However, the contribution of transition to this state in the total S factor is relatively small and the difference between  GSM-CC(1) and GSM-CC(2) results is also small. %Moreover, the experimental data for this transition disagree~\cite{Imbriani2005,chen202414npg}.
%and the calculated S factor agrees closely with the experimental data.
The calculated S factor for the transition to the first excited state $1/2^+_1$ exceeds the experimental data in the non-resonant energy region. The partial S factors calculated for transitions to other states are consistent with the measured values.

The astrophysical factor at zero energy S(0) plays a crucial role in nuclear astrophysics. In the Sun, temperatures in the solar core range from 0.020 to 0.035 MeV~\cite{ATOMKI14N}.
As shown in Fig.~\ref{fig-2}(a), values of the S factor at $E_{\mathrm{c.m.}} \geq 0$ are nearly constant and match the zero-energy S factor. Therefore, the comparison of calculated and experimental S factors in Fig.~\ref{fig-3} is done at $E_{\mathrm{c.m.}} = 0$. The experimental astrophysical factors in this figure are derived by extrapolating the measured cross sections to zero energy using the R-matrix approach.

It can be seen that the GSM-CC(1) and GSM-CC(2) calculations give an astrophysical factor S(0) that is larger than the value recommended by SF-III~\cite{SF-III} and previous experimental data. However, the GSM-CC(2) calculation yields the S(0) factor (1.996 keV b) which is compatible within error bars with a recent measurement at the HINEG facility (1.92 $\pm$ 0.08 keV b)~\cite{chen202414npg}.

It is worth noticing that GSM-CC(1) and GSM-CC(2) calculations, which yield quite similar energy spectrum of $^{15}$O, predict significantly different S factor at the zero energy. They are: S(0)=2.263 keV b and S(0)=1.996 keV b for GSM-CC(1) and GSM-CC(2), respectively. The partial S factors at the zero energy are shown in Table~\ref{tab:S}.
Analyzing the partial S factors shown in Fig. \ref{fig-2} and Table~\ref{tab:S}, it can be seen that despite the main difference between the GSM-CC(1) and GSM-CC(2) spectra, which corresponds to the energy level 5/2$^+_1$, the difference between the partial S(0) factors in these two calculations is visible in the transitions to several final states: 1/2$^-_1$, 1/2$^+_1$ and 5/2$^+_1$. This indicates a high sensitivity of the calculated astrophysical factor for the reaction $^{14}$N(p,$\gamma)^{15}$O at the zero energy to even small changes in the wave functions of several states of $^{15}$O. In this respect, energies of $5/2^+_1$ and $5/2^+_2$ levels are the largest source of uncertainties in our calculations.

%To further investigate the effect of our theoretical calculation of  S factors on solar composition problem, one
%can derived the C and N abundances relative to the hydrogen (N$_{CN}$) at the photosphere of the sun
%with the method described in Ref.~\cite{14NNeutrinos2022}. Based on the B23-GS98 analysis together with the recent global analysis of neutrino measurements, with the calculated S factor from GSM-CC(1), N$_{CN}$ is 3.61$\times10^{-4}$, and its value is 4.15$\times10^{-4}$ with GSM-CC(2) S factor. N$_{CN}$ from predictions still has large uncertainties. Based on spectroscopy of the photosphere, N$_{CN}$=4.143$\times10^{-4}$ from the GS98 compilations~\cite{Grevesse1998}.
%N$_{CN}$=5.81$_{-0.94}^{+1.22}\times10^{-4}$ given from the solar neutrino measurement~\cite{Basilico23}, N$_{CN}$=4.45$_{-0.61}^{+0.69}\times10^{-4}$ derived from the latest measurement of $^{14}$N(p,$\gamma)^{15}$O reaction cross section at Hefei.

To further investigate impact of the astrophysical S factor calculated in GSM-CC on the composition of the Sun, we determine the abundances of C and N nuclei relative to hydrogen ($N_{\mathrm{CN}}$) in the Sun's photosphere using the method described in Ref.~\cite{14NNeutrinos2022}. Based on the B23-GS98 analysis combined with recent global neutrino measurements and the GSM-CC astrophysical factor, the derived $N_{\mathrm{CN}}$ values are $3.61\times10^{-4}$
and $4.15\times10^{-4}$ for GSM-CC(1) and GSM-CC(2), respectively.

Current predictions of $N_{\mathrm{CN}}$ still exhibit significant uncertainties. The measurements based on spectroscopy of the photosphere yield $N_{\mathrm{CN}} = 4.143\times10^{-4}$ according to the GS98 compilations~\cite{Grevesse1998}. For comparison, solar neutrino measurements give $N_{\mathrm{CN}} = 5.81_{-0.94}^{+1.22}\times10^{-4}$~\cite{Basilico23}, while the latest measurement of the cross section of the $^{14}$N(p,$\gamma)^{15}$O reaction at HINEG gives $N_{\mathrm{CN}} = 4.45_{-0.61}^{+0.69}\times10^{-4}$~\cite{chen202414npg}. GSM-CC(1) calculation seems to underestimate the $N_{\mathrm{CN}}$ compared to both the spectroscopy measurement and neutrino data, whereas GSM-CC(2) demonstrates improved consistency with the data.

%The S factors derived from experimental data are the extrapolation results from the R-matrix method.
%\begin{figure}[htbp]
%	\includegraphics[width=1.0\linewidth]{diff1I2-3I2+.pdf}
%	\caption{The differential S factors for the proton capture of $^{14}$N to the ground state and the 6.79 MeV excited state of $^{15}$O  are shown in Panel (a) and (b) respectively. The experimental data are listed for comparisons~\cite{chen202414npg,Li201614N,Wagner201814N,Frentz202214N,Runkle200514N,Imbriani200514N}. The data reported in Refs.~\cite{chen202414npg,Wagner201814N,Runkle200514N,Imbriani200514N} are given as angle integrated, and are converted as the   differential S factor with the scaling factor of 4$\pi$. }
%	\label{fig-diff}
%\end{figure}

%When measuring the reaction cross section, the detecter was placed at 0$^{\circ}$ and 55$^{\circ}$, as the experiments done by Runkle \texit{et al.}~\cite{Runkle200514N}, Imbriani \texit{et al.}~\cite{Imbriani200514N}, Wagner \texit{et al.}~\cite{Wagner201814N} and Frentz el al.~\cite{Frentz202214N}. Especially, the differential S factors at 55$^{\circ}$ were reported in Ref.~\cite{Frentz202214N}. In Fig.~\ref{fig-diff}, the differential S factor for the transition to the ground state and 6.79 MeV excited state are shown. The variations of calculated differential S factor for different angles are small for the transition to ground state of $^{15}$O, and the variations are larger for the transition to the 6.79 MeV state. In general, the calculation results are consistent with the experimental data.

\vskip 0.2 truecm
\textit{Conclusions --}
The $^{14}$N(p,$\gamma)^{15}$O reaction cross section has been studied here using the microscopic GSM-CC approach which provides a unified description of spectra and reaction cross-sections. As the slowest process in the CNO cycle, this reaction is crucial for addressing the solar metallicity problem. Notably, recent direct measurements \cite{chen202414npg} report the value of S factor that is significantly higher than the currently recommended value in the SF-III compilation~\cite{SF-III}. To demonstrate theoretical uncertainties, we implemented two variants of the GSM-CC calculation: the one from fitting the energies of low-lying states of $^{14}$N and $^{15}$O, and the other one with a slight increase of the depth of the Woods-Saxon potential
for the $sd$ shells. After adjusting the corrective factors in the GSM-CC calculation, both approaches yield a similar spectroscopy of $^{15}$O and the calculated astrophysical factors show good agreement with experimental data. The S factors at zero energy calculated in GSM-CC are higher than most previous experimental extrapolations. The GSM-CC(2) calculation reduces S(0) as compared with GSM-CC(1), and is compatible with the value provided in the recent measurement at HINEG~\cite{chen202414npg}.

The microscopic calculations presented in this work yield C and N abundances $N_{\mathrm{CN}} \sim 3.61$--$4.15\times10^{-4}$, consistent within $1\sigma$ confidence level with recent cross-section measurements at HINEG facility~\cite{chen202414npg}, but significantly lower than $N_{\mathrm{CN}}$ values derived from the solar neutrino measurements~\cite{Basilico23}.

%From their data for S factor, high-metallicity composition of the Sun is favored, but still having the possibility for its low-metallicity compostion.

%thus, the high-metallicity composition is supported from our theoretical calculation.

\vskip 0.2 truecm
\textit{Acknowledgements --} This work has been supported by the National Natural Science Foundation of China under Grant Nos.~12275081, 12175281, 11605054, 12347106, and U2067205. G.X.D.~and X.B.W.~contributed equally to this work.

%\bibliography{14n15o}

\begin{thebibliography}{55}%
\makeatletter
\providecommand \@ifxundefined [1]{%
 \@ifx{#1\undefined}
}%
\providecommand \@ifnum [1]{%
 \ifnum #1\expandafter \@firstoftwo
 \else \expandafter \@secondoftwo
 \fi
}%
\providecommand \@ifx [1]{%
 \ifx #1\expandafter \@firstoftwo
 \else \expandafter \@secondoftwo
 \fi
}%
\providecommand \natexlab [1]{#1}%
\providecommand \enquote  [1]{``#1''}%
\providecommand \bibnamefont  [1]{#1}%
\providecommand \bibfnamefont [1]{#1}%
\providecommand \citenamefont [1]{#1}%
\providecommand \href@noop [0]{\@secondoftwo}%
\providecommand \href [0]{\begingroup \@sanitize@url \@href}%
\providecommand \@href[1]{\@@startlink{#1}\@@href}%
\providecommand \@@href[1]{\endgroup#1\@@endlink}%
\providecommand \@sanitize@url [0]{\catcode `\\12\catcode `\$12\catcode
  `\&12\catcode `\#12\catcode `\^12\catcode `\_12\catcode `\%12\relax}%
\providecommand \@@startlink[1]{}%
\providecommand \@@endlink[0]{}%
\providecommand \url  [0]{\begingroup\@sanitize@url \@url }%
\providecommand \@url [1]{\endgroup\@href {#1}{\urlprefix }}%
\providecommand \urlprefix  [0]{URL }%
\providecommand \Eprint [0]{\href }%
\providecommand \doibase [0]{http://dx.doi.org/}%
\providecommand \selectlanguage [0]{\@gobble}%
\providecommand \bibinfo  [0]{\@secondoftwo}%
\providecommand \bibfield  [0]{\@secondoftwo}%
\providecommand \translation [1]{[#1]}%
\providecommand \BibitemOpen [0]{}%
\providecommand \bibitemStop [0]{}%
\providecommand \bibitemNoStop [0]{.\EOS\space}%
\providecommand \EOS [0]{\spacefactor3000\relax}%
\providecommand \BibitemShut  [1]{\csname bibitem#1\endcsname}%
\let\auto@bib@innerbib\@empty
%</preamble>
\bibitem [{\citenamefont {Bahcall}\ and\ \citenamefont
  {Pinsonneault}(2004)}]{Bahcall04}%
  \BibitemOpen
  \bibfield  {author} {\bibinfo {author} {\bibfnamefont {J.~N.}\ \bibnamefont
  {Bahcall}}\ and\ \bibinfo {author} {\bibfnamefont {M.~H.}\ \bibnamefont
  {Pinsonneault}},\ }\href {\doibase 10.1103/PhysRevLett.92.121301} {\bibfield
  {journal} {\bibinfo  {journal} {Phys. Rev. Lett.}\ }\textbf {\bibinfo
  {volume} {92}},\ \bibinfo {pages} {121301} (\bibinfo {year}
  {2004})}\BibitemShut {NoStop}%
\bibitem [{\citenamefont {Bahcall}\ \emph {et~al.}(2006)\citenamefont
  {Bahcall}, \citenamefont {Serenelli},\ and\ \citenamefont
  {Basu}}]{Bahcall_2006}%
  \BibitemOpen
  \bibfield  {author} {\bibinfo {author} {\bibfnamefont {J.~N.}\ \bibnamefont
  {Bahcall}}, \bibinfo {author} {\bibfnamefont {A.~M.}\ \bibnamefont
  {Serenelli}}, \ and\ \bibinfo {author} {\bibfnamefont {S.}~\bibnamefont
  {Basu}},\ }\href {\doibase 10.1086/504043} {\bibfield  {journal} {\bibinfo
  {journal} {The Astrophysical Journal Supplement Series}\ }\textbf {\bibinfo
  {volume} {165}},\ \bibinfo {pages} {400} (\bibinfo {year}
  {2006})}\BibitemShut {NoStop}%
\bibitem [{\citenamefont {Bahcall}\ \emph {et~al.}(2005)\citenamefont
  {Bahcall}, \citenamefont {Basu}, \citenamefont {Pinsonneault},\ and\
  \citenamefont {Serenelli}}]{Bahcall_2005}%
  \BibitemOpen
  \bibfield  {author} {\bibinfo {author} {\bibfnamefont {J.~N.}\ \bibnamefont
  {Bahcall}}, \bibinfo {author} {\bibfnamefont {S.}~\bibnamefont {Basu}},
  \bibinfo {author} {\bibfnamefont {M.}~\bibnamefont {Pinsonneault}}, \ and\
  \bibinfo {author} {\bibfnamefont {A.~M.}\ \bibnamefont {Serenelli}},\ }\href
  {\doibase 10.1086/426070} {\bibfield  {journal} {\bibinfo  {journal} {The
  Astrophysical Journal}\ }\textbf {\bibinfo {volume} {618}},\ \bibinfo {pages}
  {1049} (\bibinfo {year} {2005})}\BibitemShut {NoStop}%
\bibitem [{Note1()}]{Note1}%
  \BibitemOpen
  \bibinfo {note} {The main improvement of the technique stems from improved
  atomic and molecular transition probabilities, the development of 3D
  hydrodynamical solar model atmosphere, and non-local thermodynamic
  equilibrium (LTE) in the spectral line formation~\cite
  {Asplund09,Caffau2011,Asplund2021}.}\BibitemShut {Stop}%
\bibitem [{\citenamefont {Asplund}\ \emph {et~al.}(2009)\citenamefont
  {Asplund}, \citenamefont {Grevesse}, \citenamefont {Sauval},\ and\
  \citenamefont {Scott}}]{Asplund09}%
  \BibitemOpen
  \bibfield  {author} {\bibinfo {author} {\bibfnamefont {M.}~\bibnamefont
  {Asplund}}, \bibinfo {author} {\bibfnamefont {N.}~\bibnamefont {Grevesse}},
  \bibinfo {author} {\bibfnamefont {A.~J.}\ \bibnamefont {Sauval}}, \ and\
  \bibinfo {author} {\bibfnamefont {P.}~\bibnamefont {Scott}},\ }\href
  {\doibase https://doi.org/10.1146/annurev.astro.46.060407.145222} {\bibfield
  {journal} {\bibinfo  {journal} {Annu. Rev. Astron. Astrophys.}\ }\textbf
  {\bibinfo {volume} {47}},\ \bibinfo {pages} {481} (\bibinfo {year}
  {2009})}\BibitemShut {NoStop}%
\bibitem [{\citenamefont {{Caffau}}\ \emph {et~al.}(2011)\citenamefont
  {{Caffau}}, \citenamefont {{Ludwig}}, \citenamefont {{Steffen}},
  \citenamefont {{Freytag}},\ and\ \citenamefont {{Bonifacio}}}]{Caffau2011}%
  \BibitemOpen
  \bibfield  {author} {\bibinfo {author} {\bibfnamefont {E.}~\bibnamefont
  {{Caffau}}}, \bibinfo {author} {\bibfnamefont {H.~G.}\ \bibnamefont
  {{Ludwig}}}, \bibinfo {author} {\bibfnamefont {M.}~\bibnamefont {{Steffen}}},
  \bibinfo {author} {\bibfnamefont {B.}~\bibnamefont {{Freytag}}}, \ and\
  \bibinfo {author} {\bibfnamefont {P.}~\bibnamefont {{Bonifacio}}},\ }\href
  {\doibase 10.1007/s11207-010-9541-4} {\bibfield  {journal} {\bibinfo
  {journal} {Sol. Phys.}\ }\textbf {\bibinfo {volume} {268}},\ \bibinfo {pages}
  {255} (\bibinfo {year} {2011})},\ \Eprint {http://arxiv.org/abs/1003.1190}
  {arXiv:1003.1190 [astro-ph.SR]} \BibitemShut {NoStop}%
\bibitem [{\citenamefont {{Asplund}}\ \emph {et~al.}(2021)\citenamefont
  {{Asplund}}, \citenamefont {{Amarsi}},\ and\ \citenamefont
  {{Grevesse}}}]{Asplund2021}%
  \BibitemOpen
  \bibfield  {author} {\bibinfo {author} {\bibfnamefont {M.}~\bibnamefont
  {{Asplund}}}, \bibinfo {author} {\bibfnamefont {A.~M.}\ \bibnamefont
  {{Amarsi}}}, \ and\ \bibinfo {author} {\bibfnamefont {N.}~\bibnamefont
  {{Grevesse}}},\ }\href {\doibase 10.1051/0004-6361/202140445} {\bibfield
  {journal} {\bibinfo  {journal} {Astron. Astrophys.}\ }\textbf {\bibinfo
  {volume} {653}},\ \bibinfo {eid} {A141} (\bibinfo {year} {2021})},\ \Eprint
  {http://arxiv.org/abs/2105.01661} {arXiv:2105.01661 [astro-ph.SR]}
  \BibitemShut {NoStop}%
\bibitem [{\citenamefont {{Grevesse}}\ and\ \citenamefont
  {{Sauval}}(1998)}]{Grevesse1998}%
  \BibitemOpen
  \bibfield  {author} {\bibinfo {author} {\bibfnamefont {N.}~\bibnamefont
  {{Grevesse}}}\ and\ \bibinfo {author} {\bibfnamefont {A.~J.}\ \bibnamefont
  {{Sauval}}},\ }\href {\doibase 10.1023/A:1005161325181} {\bibfield  {journal}
  {\bibinfo  {journal} {Space Sci. Rev.}\ }\textbf {\bibinfo {volume} {85}},\
  \bibinfo {pages} {161} (\bibinfo {year} {1998})}\BibitemShut {NoStop}%
\bibitem [{\citenamefont {Haxton}\ and\ \citenamefont
  {Serenelli}(2008)}]{Haxton_2008}%
  \BibitemOpen
  \bibfield  {author} {\bibinfo {author} {\bibfnamefont {W.~C.}\ \bibnamefont
  {Haxton}}\ and\ \bibinfo {author} {\bibfnamefont {A.~M.}\ \bibnamefont
  {Serenelli}},\ }\href {\doibase 10.1086/591787} {\bibfield  {journal}
  {\bibinfo  {journal} {The Astrophysical Journal}\ }\textbf {\bibinfo {volume}
  {687}},\ \bibinfo {pages} {678} (\bibinfo {year} {2008})}\BibitemShut
  {NoStop}%
\bibitem [{\citenamefont {{Wiescher}}(2018)}]{Wiescher18}%
  \BibitemOpen
  \bibfield  {author} {\bibinfo {author} {\bibfnamefont {M.}~\bibnamefont
  {{Wiescher}}},\ }\href {\doibase 10.1007/s00016-018-0216-0} {\bibfield
  {journal} {\bibinfo  {journal} {Phys. Perspect.}\ }\textbf {\bibinfo {volume}
  {20}},\ \bibinfo {pages} {124} (\bibinfo {year} {2018})}\BibitemShut
  {NoStop}%
\bibitem [{\citenamefont {Agostini}\ and\ \citenamefont
  {et~al.}(2020)}]{14NNeutrinos2020}%
  \BibitemOpen
  \bibfield  {author} {\bibinfo {author} {\bibfnamefont {M.}~\bibnamefont
  {Agostini}}\ and\ \bibinfo {author} {\bibnamefont {et~al.}} (\bibinfo
  {collaboration} {Borexino Collaboration}),\ }\href {\doibase
  10.1038/s41586-020-2934-0} {\bibfield  {journal} {\bibinfo  {journal} {Nature
  (London)}\ }\textbf {\bibinfo {volume} {587}},\ \bibinfo {pages} {577}
  (\bibinfo {year} {2020})}\BibitemShut {NoStop}%
\bibitem [{\citenamefont {Appel}\ \emph {et~al.}(2022)\citenamefont {Appel},
  \citenamefont {Bagdasarian}, \citenamefont {Basilico}, \citenamefont
  {Bellini}, \citenamefont {Benziger}, \citenamefont {Biondi}, \citenamefont
  {Caccianiga}, \citenamefont {Calaprice}, \citenamefont {Caminata},
  \citenamefont {Cavalcante}, \citenamefont {Chepurnov}, \citenamefont
  {D'Angelo}, \citenamefont {Derbin}, \citenamefont {Di~Giacinto},
  \citenamefont {Di~Marcello}, \citenamefont {Ding}, \citenamefont
  {Di~Ludovico}, \citenamefont {Di~Noto}, \citenamefont {Drachnev},
  \citenamefont {Franco}, \citenamefont {Galbiati}, \citenamefont {Ghiano},
  \citenamefont {Giammarchi}, \citenamefont {Goretti}, \citenamefont
  {G\"ottel}, \citenamefont {Gromov}, \citenamefont {Guffanti}, \citenamefont
  {Ianni}, \citenamefont {Ianni}, \citenamefont {Jany}, \citenamefont
  {Kobychev}, \citenamefont {Korga}, \citenamefont {Kumaran}, \citenamefont
  {Laubenstein}, \citenamefont {Litvinovich}, \citenamefont {Lombardi},
  \citenamefont {Lomskaya}, \citenamefont {Ludhova}, \citenamefont
  {Lukyanchenko}, \citenamefont {Machulin}, \citenamefont {Martyn},
  \citenamefont {Meroni}, \citenamefont {Miramonti}, \citenamefont {Misiaszek},
  \citenamefont {Muratova}, \citenamefont {Nugmanov}, \citenamefont {Oberauer},
  \citenamefont {Orekhov}, \citenamefont {Ortica}, \citenamefont {Pallavicini},
  \citenamefont {Papp}, \citenamefont {Pelicci}, \citenamefont {Penek},
  \citenamefont {Pietrofaccia}, \citenamefont {Pilipenko}, \citenamefont
  {Pocar}, \citenamefont {Raikov}, \citenamefont {Ranalli}, \citenamefont
  {Ranucci}, \citenamefont {Razeto}, \citenamefont {Re}, \citenamefont
  {Redchuk}, \citenamefont {Rossi}, \citenamefont {Sch\"onert}, \citenamefont
  {Semenov}, \citenamefont {Settanta}, \citenamefont {Skorokhvatov},
  \citenamefont {Singhal}, \citenamefont {Smirnov}, \citenamefont {Sotnikov},
  \citenamefont {Tartaglia}, \citenamefont {Testera}, \citenamefont {Unzhakov},
  \citenamefont {Villante}, \citenamefont {Vishneva}, \citenamefont {Vogelaar},
  \citenamefont {von Feilitzsch}, \citenamefont {Wojcik}, \citenamefont {Wurm},
  \citenamefont {Zavatarelli}, \citenamefont {Zuber},\ and\ \citenamefont
  {Zuzel}}]{14NNeutrinos2022}%
  \BibitemOpen
  \bibfield  {author} {\bibinfo {author} {\bibfnamefont {S.}~\bibnamefont
  {Appel}}, \bibinfo {author} {\bibfnamefont {Z.}~\bibnamefont {Bagdasarian}},
  \bibinfo {author} {\bibfnamefont {D.}~\bibnamefont {Basilico}}, \bibinfo
  {author} {\bibfnamefont {G.}~\bibnamefont {Bellini}}, \bibinfo {author}
  {\bibfnamefont {J.}~\bibnamefont {Benziger}}, \bibinfo {author}
  {\bibfnamefont {R.}~\bibnamefont {Biondi}}, \bibinfo {author} {\bibfnamefont
  {B.}~\bibnamefont {Caccianiga}}, \bibinfo {author} {\bibfnamefont
  {F.}~\bibnamefont {Calaprice}}, \bibinfo {author} {\bibfnamefont
  {A.}~\bibnamefont {Caminata}}, \bibinfo {author} {\bibfnamefont
  {P.}~\bibnamefont {Cavalcante}}, \bibinfo {author} {\bibfnamefont
  {A.}~\bibnamefont {Chepurnov}}, \bibinfo {author} {\bibfnamefont
  {D.}~\bibnamefont {D'Angelo}}, \bibinfo {author} {\bibfnamefont
  {A.}~\bibnamefont {Derbin}}, \bibinfo {author} {\bibfnamefont
  {A.}~\bibnamefont {Di~Giacinto}}, \bibinfo {author} {\bibfnamefont
  {V.}~\bibnamefont {Di~Marcello}}, \bibinfo {author} {\bibfnamefont {X.~F.}\
  \bibnamefont {Ding}}, \bibinfo {author} {\bibfnamefont {A.}~\bibnamefont
  {Di~Ludovico}}, \bibinfo {author} {\bibfnamefont {L.}~\bibnamefont
  {Di~Noto}}, \bibinfo {author} {\bibfnamefont {I.}~\bibnamefont {Drachnev}},
  \bibinfo {author} {\bibfnamefont {D.}~\bibnamefont {Franco}}, \bibinfo
  {author} {\bibfnamefont {C.}~\bibnamefont {Galbiati}}, \bibinfo {author}
  {\bibfnamefont {C.}~\bibnamefont {Ghiano}}, \bibinfo {author} {\bibfnamefont
  {M.}~\bibnamefont {Giammarchi}}, \bibinfo {author} {\bibfnamefont
  {A.}~\bibnamefont {Goretti}}, \bibinfo {author} {\bibfnamefont {A.~S.}\
  \bibnamefont {G\"ottel}}, \bibinfo {author} {\bibfnamefont {M.}~\bibnamefont
  {Gromov}}, \bibinfo {author} {\bibfnamefont {D.}~\bibnamefont {Guffanti}},
  \bibinfo {author} {\bibfnamefont {A.}~\bibnamefont {Ianni}}, \bibinfo
  {author} {\bibfnamefont {A.}~\bibnamefont {Ianni}}, \bibinfo {author}
  {\bibfnamefont {A.}~\bibnamefont {Jany}}, \bibinfo {author} {\bibfnamefont
  {V.}~\bibnamefont {Kobychev}}, \bibinfo {author} {\bibfnamefont
  {G.}~\bibnamefont {Korga}}, \bibinfo {author} {\bibfnamefont
  {S.}~\bibnamefont {Kumaran}}, \bibinfo {author} {\bibfnamefont
  {M.}~\bibnamefont {Laubenstein}}, \bibinfo {author} {\bibfnamefont
  {E.}~\bibnamefont {Litvinovich}}, \bibinfo {author} {\bibfnamefont
  {P.}~\bibnamefont {Lombardi}}, \bibinfo {author} {\bibfnamefont
  {I.}~\bibnamefont {Lomskaya}}, \bibinfo {author} {\bibfnamefont
  {L.}~\bibnamefont {Ludhova}}, \bibinfo {author} {\bibfnamefont
  {G.}~\bibnamefont {Lukyanchenko}}, \bibinfo {author} {\bibfnamefont
  {I.}~\bibnamefont {Machulin}}, \bibinfo {author} {\bibfnamefont
  {J.}~\bibnamefont {Martyn}}, \bibinfo {author} {\bibfnamefont
  {E.}~\bibnamefont {Meroni}}, \bibinfo {author} {\bibfnamefont
  {L.}~\bibnamefont {Miramonti}}, \bibinfo {author} {\bibfnamefont
  {M.}~\bibnamefont {Misiaszek}}, \bibinfo {author} {\bibfnamefont
  {V.}~\bibnamefont {Muratova}}, \bibinfo {author} {\bibfnamefont
  {R.}~\bibnamefont {Nugmanov}}, \bibinfo {author} {\bibfnamefont
  {L.}~\bibnamefont {Oberauer}}, \bibinfo {author} {\bibfnamefont
  {V.}~\bibnamefont {Orekhov}}, \bibinfo {author} {\bibfnamefont
  {F.}~\bibnamefont {Ortica}}, \bibinfo {author} {\bibfnamefont
  {M.}~\bibnamefont {Pallavicini}}, \bibinfo {author} {\bibfnamefont
  {L.}~\bibnamefont {Papp}}, \bibinfo {author} {\bibfnamefont {L.}~\bibnamefont
  {Pelicci}}, \bibinfo {author} {\bibfnamefont {O.}~\bibnamefont {Penek}},
  \bibinfo {author} {\bibfnamefont {L.}~\bibnamefont {Pietrofaccia}}, \bibinfo
  {author} {\bibfnamefont {N.}~\bibnamefont {Pilipenko}}, \bibinfo {author}
  {\bibfnamefont {A.}~\bibnamefont {Pocar}}, \bibinfo {author} {\bibfnamefont
  {G.}~\bibnamefont {Raikov}}, \bibinfo {author} {\bibfnamefont {M.~T.}\
  \bibnamefont {Ranalli}}, \bibinfo {author} {\bibfnamefont {G.}~\bibnamefont
  {Ranucci}}, \bibinfo {author} {\bibfnamefont {A.}~\bibnamefont {Razeto}},
  \bibinfo {author} {\bibfnamefont {A.}~\bibnamefont {Re}}, \bibinfo {author}
  {\bibfnamefont {M.}~\bibnamefont {Redchuk}}, \bibinfo {author} {\bibfnamefont
  {N.}~\bibnamefont {Rossi}}, \bibinfo {author} {\bibfnamefont
  {S.}~\bibnamefont {Sch\"onert}}, \bibinfo {author} {\bibfnamefont
  {D.}~\bibnamefont {Semenov}}, \bibinfo {author} {\bibfnamefont
  {G.}~\bibnamefont {Settanta}}, \bibinfo {author} {\bibfnamefont
  {M.}~\bibnamefont {Skorokhvatov}}, \bibinfo {author} {\bibfnamefont
  {A.}~\bibnamefont {Singhal}}, \bibinfo {author} {\bibfnamefont
  {O.}~\bibnamefont {Smirnov}}, \bibinfo {author} {\bibfnamefont
  {A.}~\bibnamefont {Sotnikov}}, \bibinfo {author} {\bibfnamefont
  {R.}~\bibnamefont {Tartaglia}}, \bibinfo {author} {\bibfnamefont
  {G.}~\bibnamefont {Testera}}, \bibinfo {author} {\bibfnamefont
  {E.}~\bibnamefont {Unzhakov}}, \bibinfo {author} {\bibfnamefont {F.~L.}\
  \bibnamefont {Villante}}, \bibinfo {author} {\bibfnamefont {A.}~\bibnamefont
  {Vishneva}}, \bibinfo {author} {\bibfnamefont {R.~B.}\ \bibnamefont
  {Vogelaar}}, \bibinfo {author} {\bibfnamefont {F.}~\bibnamefont {von
  Feilitzsch}}, \bibinfo {author} {\bibfnamefont {M.}~\bibnamefont {Wojcik}},
  \bibinfo {author} {\bibfnamefont {M.}~\bibnamefont {Wurm}}, \bibinfo {author}
  {\bibfnamefont {S.}~\bibnamefont {Zavatarelli}}, \bibinfo {author}
  {\bibfnamefont {K.}~\bibnamefont {Zuber}}, \ and\ \bibinfo {author}
  {\bibfnamefont {G.}~\bibnamefont {Zuzel}} (\bibinfo {collaboration} {Borexino
  Collaboration}),\ }\href {\doibase 10.1103/PhysRevLett.129.252701} {\bibfield
   {journal} {\bibinfo  {journal} {Phys. Rev. Lett.}\ }\textbf {\bibinfo
  {volume} {129}},\ \bibinfo {pages} {252701} (\bibinfo {year}
  {2022})}\BibitemShut {NoStop}%
\bibitem [{\citenamefont {Formicola}\ \emph {et~al.}(2004)\citenamefont
  {Formicola}, \citenamefont {Imbriani}, \citenamefont {Costantini},
  \citenamefont {Angulo}, \citenamefont {Bemmerer}, \citenamefont {Bonetti},
  \citenamefont {Broggini}, \citenamefont {Corvisiero}, \citenamefont {Cruz},
  \citenamefont {Descouvemont}, \citenamefont {F¨¹l?p}, \citenamefont
  {Gervino}, \citenamefont {Guglielmetti}, \citenamefont {Gustavino},
  \citenamefont {Gy¨¹rky}, \citenamefont {Jesus}, \citenamefont {Junker},
  \citenamefont {Lemut}, \citenamefont {Menegazzo}, \citenamefont {Prati},
  \citenamefont {Roca}, \citenamefont {Rolfs}, \citenamefont {Romano},
  \citenamefont {{Rossi Alvarez}}, \citenamefont {Sch¨¹mann}, \citenamefont
  {Somorjai}, \citenamefont {Straniero}, \citenamefont {Strieder},
  \citenamefont {Terrasi}, \citenamefont {Trautvetter}, \citenamefont
  {Vomiero},\ and\ \citenamefont {Zavatarelli}}]{FORMICOLA200461}%
  \BibitemOpen
  \bibfield  {author} {\bibinfo {author} {\bibfnamefont {A.}~\bibnamefont
  {Formicola}}, \bibinfo {author} {\bibfnamefont {G.}~\bibnamefont {Imbriani}},
  \bibinfo {author} {\bibfnamefont {H.}~\bibnamefont {Costantini}}, \bibinfo
  {author} {\bibfnamefont {C.}~\bibnamefont {Angulo}}, \bibinfo {author}
  {\bibfnamefont {D.}~\bibnamefont {Bemmerer}}, \bibinfo {author}
  {\bibfnamefont {R.}~\bibnamefont {Bonetti}}, \bibinfo {author} {\bibfnamefont
  {C.}~\bibnamefont {Broggini}}, \bibinfo {author} {\bibfnamefont
  {P.}~\bibnamefont {Corvisiero}}, \bibinfo {author} {\bibfnamefont
  {J.}~\bibnamefont {Cruz}}, \bibinfo {author} {\bibfnamefont {P.}~\bibnamefont
  {Descouvemont}}, \bibinfo {author} {\bibfnamefont {Z.}~\bibnamefont
  {F¨¹l?p}}, \bibinfo {author} {\bibfnamefont {G.}~\bibnamefont {Gervino}},
  \bibinfo {author} {\bibfnamefont {A.}~\bibnamefont {Guglielmetti}}, \bibinfo
  {author} {\bibfnamefont {C.}~\bibnamefont {Gustavino}}, \bibinfo {author}
  {\bibfnamefont {G.}~\bibnamefont {Gy¨¹rky}}, \bibinfo {author} {\bibfnamefont
  {A.}~\bibnamefont {Jesus}}, \bibinfo {author} {\bibfnamefont
  {M.}~\bibnamefont {Junker}}, \bibinfo {author} {\bibfnamefont
  {A.}~\bibnamefont {Lemut}}, \bibinfo {author} {\bibfnamefont
  {R.}~\bibnamefont {Menegazzo}}, \bibinfo {author} {\bibfnamefont
  {P.}~\bibnamefont {Prati}}, \bibinfo {author} {\bibfnamefont
  {V.}~\bibnamefont {Roca}}, \bibinfo {author} {\bibfnamefont {C.}~\bibnamefont
  {Rolfs}}, \bibinfo {author} {\bibfnamefont {M.}~\bibnamefont {Romano}},
  \bibinfo {author} {\bibfnamefont {C.}~\bibnamefont {{Rossi Alvarez}}},
  \bibinfo {author} {\bibfnamefont {F.}~\bibnamefont {Sch¨¹mann}}, \bibinfo
  {author} {\bibfnamefont {E.}~\bibnamefont {Somorjai}}, \bibinfo {author}
  {\bibfnamefont {O.}~\bibnamefont {Straniero}}, \bibinfo {author}
  {\bibfnamefont {F.}~\bibnamefont {Strieder}}, \bibinfo {author}
  {\bibfnamefont {F.}~\bibnamefont {Terrasi}}, \bibinfo {author} {\bibfnamefont
  {H.}~\bibnamefont {Trautvetter}}, \bibinfo {author} {\bibfnamefont
  {A.}~\bibnamefont {Vomiero}}, \ and\ \bibinfo {author} {\bibfnamefont
  {S.}~\bibnamefont {Zavatarelli}},\ }\href {\doibase
  https://doi.org/10.1016/j.physletb.2004.03.092} {\bibfield  {journal}
  {\bibinfo  {journal} {Physics Letters B}\ }\textbf {\bibinfo {volume}
  {591}},\ \bibinfo {pages} {61} (\bibinfo {year} {2004})}\BibitemShut
  {NoStop}%
\bibitem [{\citenamefont {Imbriani}\ \emph {et~al.}(2005)\citenamefont
  {Imbriani}, \citenamefont {Costantini}, \citenamefont {Formicola},\ and\
  \citenamefont {et~al.}}]{Imbriani200514N}%
  \BibitemOpen
  \bibfield  {author} {\bibinfo {author} {\bibfnamefont {G.}~\bibnamefont
  {Imbriani}}, \bibinfo {author} {\bibfnamefont {H.}~\bibnamefont
  {Costantini}}, \bibinfo {author} {\bibfnamefont {A.}~\bibnamefont
  {Formicola}}, \ and\ \bibinfo {author} {\bibnamefont {et~al.}},\ }\href@noop
  {} {\bibfield  {journal} {\bibinfo  {journal} {Eur. Phys. J. A}\ }\textbf
  {\bibinfo {volume} {25}},\ \bibinfo {pages} {455} (\bibinfo {year}
  {2005})}\BibitemShut {NoStop}%
\bibitem [{\citenamefont {Bemmerer}\ \emph {et~al.}(2006)\citenamefont
  {Bemmerer}, \citenamefont {Confortola}, \citenamefont {Lemut}, \citenamefont
  {Bonetti}, \citenamefont {Broggini}, \citenamefont {Corvisiero},
  \citenamefont {Costantini}, \citenamefont {Cruz}, \citenamefont {Formicola},
  \citenamefont {F¨¹l?p}, \citenamefont {Gervino}, \citenamefont
  {Guglielmetti}, \citenamefont {Gustavino}, \citenamefont {Gy¨¹rky},
  \citenamefont {Imbriani}, \citenamefont {Jesus}, \citenamefont {Junker},
  \citenamefont {Limata}, \citenamefont {Menegazzo}, \citenamefont {Prati},
  \citenamefont {Roca}, \citenamefont {Rolfs}, \citenamefont {Rogalla},
  \citenamefont {Romano}, \citenamefont {Rossi-Alvarez}, \citenamefont
  {Sch¨¹mann}, \citenamefont {Somorjai}, \citenamefont {Straniero},
  \citenamefont {Strieder}, \citenamefont {Terrasi},\ and\ \citenamefont
  {Trautvetter}}]{Bemmerer200614N}%
  \BibitemOpen
  \bibfield  {author} {\bibinfo {author} {\bibfnamefont {D.}~\bibnamefont
  {Bemmerer}}, \bibinfo {author} {\bibfnamefont {F.}~\bibnamefont
  {Confortola}}, \bibinfo {author} {\bibfnamefont {A.}~\bibnamefont {Lemut}},
  \bibinfo {author} {\bibfnamefont {R.}~\bibnamefont {Bonetti}}, \bibinfo
  {author} {\bibfnamefont {C.}~\bibnamefont {Broggini}}, \bibinfo {author}
  {\bibfnamefont {P.}~\bibnamefont {Corvisiero}}, \bibinfo {author}
  {\bibfnamefont {H.}~\bibnamefont {Costantini}}, \bibinfo {author}
  {\bibfnamefont {J.}~\bibnamefont {Cruz}}, \bibinfo {author} {\bibfnamefont
  {A.}~\bibnamefont {Formicola}}, \bibinfo {author} {\bibfnamefont
  {Z.}~\bibnamefont {F¨¹l?p}}, \bibinfo {author} {\bibfnamefont
  {G.}~\bibnamefont {Gervino}}, \bibinfo {author} {\bibfnamefont
  {A.}~\bibnamefont {Guglielmetti}}, \bibinfo {author} {\bibfnamefont
  {C.}~\bibnamefont {Gustavino}}, \bibinfo {author} {\bibfnamefont
  {G.}~\bibnamefont {Gy¨¹rky}}, \bibinfo {author} {\bibfnamefont
  {G.}~\bibnamefont {Imbriani}}, \bibinfo {author} {\bibfnamefont
  {A.}~\bibnamefont {Jesus}}, \bibinfo {author} {\bibfnamefont
  {M.}~\bibnamefont {Junker}}, \bibinfo {author} {\bibfnamefont
  {B.}~\bibnamefont {Limata}}, \bibinfo {author} {\bibfnamefont
  {R.}~\bibnamefont {Menegazzo}}, \bibinfo {author} {\bibfnamefont
  {P.}~\bibnamefont {Prati}}, \bibinfo {author} {\bibfnamefont
  {V.}~\bibnamefont {Roca}}, \bibinfo {author} {\bibfnamefont {C.}~\bibnamefont
  {Rolfs}}, \bibinfo {author} {\bibfnamefont {D.}~\bibnamefont {Rogalla}},
  \bibinfo {author} {\bibfnamefont {M.}~\bibnamefont {Romano}}, \bibinfo
  {author} {\bibfnamefont {C.}~\bibnamefont {Rossi-Alvarez}}, \bibinfo {author}
  {\bibfnamefont {F.}~\bibnamefont {Sch¨¹mann}}, \bibinfo {author}
  {\bibfnamefont {E.}~\bibnamefont {Somorjai}}, \bibinfo {author}
  {\bibfnamefont {O.}~\bibnamefont {Straniero}}, \bibinfo {author}
  {\bibfnamefont {F.}~\bibnamefont {Strieder}}, \bibinfo {author}
  {\bibfnamefont {F.}~\bibnamefont {Terrasi}}, \ and\ \bibinfo {author}
  {\bibfnamefont {H.}~\bibnamefont {Trautvetter}},\ }\href {\doibase
  https://doi.org/10.1016/j.nuclphysa.2006.09.001} {\bibfield  {journal}
  {\bibinfo  {journal} {Nuclear Physics A}\ }\textbf {\bibinfo {volume}
  {779}},\ \bibinfo {pages} {297} (\bibinfo {year} {2006})}\BibitemShut
  {NoStop}%
\bibitem [{\citenamefont {Lemut}\ \emph {et~al.}(2006)\citenamefont {Lemut},
  \citenamefont {Bemmerer}, \citenamefont {Confortola}, \citenamefont
  {Bonetti}, \citenamefont {Broggini}, \citenamefont {Corvisiero},
  \citenamefont {Costantini}, \citenamefont {Cruz}, \citenamefont {Formicola},
  \citenamefont {F¨¹l?p}, \citenamefont {Gervino}, \citenamefont
  {Guglielmetti}, \citenamefont {Gustavino}, \citenamefont {Gy¨¹rky},
  \citenamefont {Imbriani}, \citenamefont {Jesus}, \citenamefont {Junker},
  \citenamefont {Limata}, \citenamefont {Menegazzo}, \citenamefont {Prati},
  \citenamefont {Roca}, \citenamefont {Rogalla}, \citenamefont {Rolfs},
  \citenamefont {Romano}, \citenamefont {{Rossi Alvarez}}, \citenamefont
  {Sch¨¹mann}, \citenamefont {Somorjai}, \citenamefont {Straniero},
  \citenamefont {Strieder}, \citenamefont {Terrasi},\ and\ \citenamefont
  {Trautvetter}}]{LEMUT2006483}%
  \BibitemOpen
  \bibfield  {author} {\bibinfo {author} {\bibfnamefont {A.}~\bibnamefont
  {Lemut}}, \bibinfo {author} {\bibfnamefont {D.}~\bibnamefont {Bemmerer}},
  \bibinfo {author} {\bibfnamefont {F.}~\bibnamefont {Confortola}}, \bibinfo
  {author} {\bibfnamefont {R.}~\bibnamefont {Bonetti}}, \bibinfo {author}
  {\bibfnamefont {C.}~\bibnamefont {Broggini}}, \bibinfo {author}
  {\bibfnamefont {P.}~\bibnamefont {Corvisiero}}, \bibinfo {author}
  {\bibfnamefont {H.}~\bibnamefont {Costantini}}, \bibinfo {author}
  {\bibfnamefont {J.}~\bibnamefont {Cruz}}, \bibinfo {author} {\bibfnamefont
  {A.}~\bibnamefont {Formicola}}, \bibinfo {author} {\bibfnamefont
  {Z.}~\bibnamefont {F¨¹l?p}}, \bibinfo {author} {\bibfnamefont
  {G.}~\bibnamefont {Gervino}}, \bibinfo {author} {\bibfnamefont
  {A.}~\bibnamefont {Guglielmetti}}, \bibinfo {author} {\bibfnamefont
  {C.}~\bibnamefont {Gustavino}}, \bibinfo {author} {\bibfnamefont
  {G.}~\bibnamefont {Gy¨¹rky}}, \bibinfo {author} {\bibfnamefont
  {G.}~\bibnamefont {Imbriani}}, \bibinfo {author} {\bibfnamefont
  {A.}~\bibnamefont {Jesus}}, \bibinfo {author} {\bibfnamefont
  {M.}~\bibnamefont {Junker}}, \bibinfo {author} {\bibfnamefont
  {B.}~\bibnamefont {Limata}}, \bibinfo {author} {\bibfnamefont
  {R.}~\bibnamefont {Menegazzo}}, \bibinfo {author} {\bibfnamefont
  {P.}~\bibnamefont {Prati}}, \bibinfo {author} {\bibfnamefont
  {V.}~\bibnamefont {Roca}}, \bibinfo {author} {\bibfnamefont {D.}~\bibnamefont
  {Rogalla}}, \bibinfo {author} {\bibfnamefont {C.}~\bibnamefont {Rolfs}},
  \bibinfo {author} {\bibfnamefont {M.}~\bibnamefont {Romano}}, \bibinfo
  {author} {\bibfnamefont {C.}~\bibnamefont {{Rossi Alvarez}}}, \bibinfo
  {author} {\bibfnamefont {F.}~\bibnamefont {Sch¨¹mann}}, \bibinfo {author}
  {\bibfnamefont {E.}~\bibnamefont {Somorjai}}, \bibinfo {author}
  {\bibfnamefont {O.}~\bibnamefont {Straniero}}, \bibinfo {author}
  {\bibfnamefont {F.}~\bibnamefont {Strieder}}, \bibinfo {author}
  {\bibfnamefont {F.}~\bibnamefont {Terrasi}}, \ and\ \bibinfo {author}
  {\bibfnamefont {H.}~\bibnamefont {Trautvetter}},\ }\href {\doibase
  https://doi.org/10.1016/j.physletb.2006.02.021} {\bibfield  {journal}
  {\bibinfo  {journal} {Physics Letters B}\ }\textbf {\bibinfo {volume}
  {634}},\ \bibinfo {pages} {483} (\bibinfo {year} {2006})}\BibitemShut
  {NoStop}%
\bibitem [{\citenamefont {Marta}\ \emph {et~al.}(2008)\citenamefont {Marta},
  \citenamefont {Formicola}, \citenamefont {Gy\"urky}, \citenamefont
  {Bemmerer}, \citenamefont {Broggini}, \citenamefont {Caciolli}, \citenamefont
  {Corvisiero}, \citenamefont {Costantini}, \citenamefont {Elekes},
  \citenamefont {F\"ul\"op}, \citenamefont {Gervino}, \citenamefont
  {Guglielmetti}, \citenamefont {Gustavino}, \citenamefont {Imbriani},
  \citenamefont {Junker}, \citenamefont {Kunz}, \citenamefont {Lemut},
  \citenamefont {Limata}, \citenamefont {Mazzocchi}, \citenamefont {Menegazzo},
  \citenamefont {Prati}, \citenamefont {Roca}, \citenamefont {Rolfs},
  \citenamefont {Romano}, \citenamefont {Alvarez}, \citenamefont {Somorjai},
  \citenamefont {Straniero}, \citenamefont {Strieder}, \citenamefont {Terrasi},
  \citenamefont {Trautvetter},\ and\ \citenamefont {Vomiero}}]{Marta0814N}%
  \BibitemOpen
  \bibfield  {author} {\bibinfo {author} {\bibfnamefont {M.}~\bibnamefont
  {Marta}}, \bibinfo {author} {\bibfnamefont {A.}~\bibnamefont {Formicola}},
  \bibinfo {author} {\bibfnamefont {G.}~\bibnamefont {Gy\"urky}}, \bibinfo
  {author} {\bibfnamefont {D.}~\bibnamefont {Bemmerer}}, \bibinfo {author}
  {\bibfnamefont {C.}~\bibnamefont {Broggini}}, \bibinfo {author}
  {\bibfnamefont {A.}~\bibnamefont {Caciolli}}, \bibinfo {author}
  {\bibfnamefont {P.}~\bibnamefont {Corvisiero}}, \bibinfo {author}
  {\bibfnamefont {H.}~\bibnamefont {Costantini}}, \bibinfo {author}
  {\bibfnamefont {Z.}~\bibnamefont {Elekes}}, \bibinfo {author} {\bibfnamefont
  {Z.}~\bibnamefont {F\"ul\"op}}, \bibinfo {author} {\bibfnamefont
  {G.}~\bibnamefont {Gervino}}, \bibinfo {author} {\bibfnamefont
  {A.}~\bibnamefont {Guglielmetti}}, \bibinfo {author} {\bibfnamefont
  {C.}~\bibnamefont {Gustavino}}, \bibinfo {author} {\bibfnamefont
  {G.}~\bibnamefont {Imbriani}}, \bibinfo {author} {\bibfnamefont
  {M.}~\bibnamefont {Junker}}, \bibinfo {author} {\bibfnamefont
  {R.}~\bibnamefont {Kunz}}, \bibinfo {author} {\bibfnamefont {A.}~\bibnamefont
  {Lemut}}, \bibinfo {author} {\bibfnamefont {B.}~\bibnamefont {Limata}},
  \bibinfo {author} {\bibfnamefont {C.}~\bibnamefont {Mazzocchi}}, \bibinfo
  {author} {\bibfnamefont {R.}~\bibnamefont {Menegazzo}}, \bibinfo {author}
  {\bibfnamefont {P.}~\bibnamefont {Prati}}, \bibinfo {author} {\bibfnamefont
  {V.}~\bibnamefont {Roca}}, \bibinfo {author} {\bibfnamefont {C.}~\bibnamefont
  {Rolfs}}, \bibinfo {author} {\bibfnamefont {M.}~\bibnamefont {Romano}},
  \bibinfo {author} {\bibfnamefont {C.~R.}\ \bibnamefont {Alvarez}}, \bibinfo
  {author} {\bibfnamefont {E.}~\bibnamefont {Somorjai}}, \bibinfo {author}
  {\bibfnamefont {O.}~\bibnamefont {Straniero}}, \bibinfo {author}
  {\bibfnamefont {F.}~\bibnamefont {Strieder}}, \bibinfo {author}
  {\bibfnamefont {F.}~\bibnamefont {Terrasi}}, \bibinfo {author} {\bibfnamefont
  {H.~P.}\ \bibnamefont {Trautvetter}}, \ and\ \bibinfo {author} {\bibfnamefont
  {A.}~\bibnamefont {Vomiero}} (\bibinfo {collaboration} {LUNA
  Collaboration}),\ }\href {\doibase 10.1103/PhysRevC.78.022802} {\bibfield
  {journal} {\bibinfo  {journal} {Phys. Rev. C}\ }\textbf {\bibinfo {volume}
  {78}},\ \bibinfo {pages} {022802} (\bibinfo {year} {2008})}\BibitemShut
  {NoStop}%
\bibitem [{\citenamefont {Runkle}\ \emph {et~al.}(2005)\citenamefont {Runkle},
  \citenamefont {Champagne}, \citenamefont {Angulo}, \citenamefont {Fox},
  \citenamefont {Iliadis}, \citenamefont {Longland},\ and\ \citenamefont
  {Pollanen}}]{Runkle200514N}%
  \BibitemOpen
  \bibfield  {author} {\bibinfo {author} {\bibfnamefont {R.~C.}\ \bibnamefont
  {Runkle}}, \bibinfo {author} {\bibfnamefont {A.~E.}\ \bibnamefont
  {Champagne}}, \bibinfo {author} {\bibfnamefont {C.}~\bibnamefont {Angulo}},
  \bibinfo {author} {\bibfnamefont {C.}~\bibnamefont {Fox}}, \bibinfo {author}
  {\bibfnamefont {C.}~\bibnamefont {Iliadis}}, \bibinfo {author} {\bibfnamefont
  {R.}~\bibnamefont {Longland}}, \ and\ \bibinfo {author} {\bibfnamefont
  {J.}~\bibnamefont {Pollanen}},\ }\href {\doibase
  10.1103/PhysRevLett.94.082503} {\bibfield  {journal} {\bibinfo  {journal}
  {Phys. Rev. Lett.}\ }\textbf {\bibinfo {volume} {94}},\ \bibinfo {pages}
  {082503} (\bibinfo {year} {2005})}\BibitemShut {NoStop}%
\bibitem [{\citenamefont {Li}\ \emph {et~al.}(2016)\citenamefont {Li},
  \citenamefont {G\"orres}, \citenamefont {deBoer}, \citenamefont {Imbriani},
  \citenamefont {Best}, \citenamefont {Kontos}, \citenamefont {LeBlanc},
  \citenamefont {Uberseder},\ and\ \citenamefont {Wiescher}}]{Li201614N}%
  \BibitemOpen
  \bibfield  {author} {\bibinfo {author} {\bibfnamefont {Q.}~\bibnamefont
  {Li}}, \bibinfo {author} {\bibfnamefont {J.}~\bibnamefont {G\"orres}},
  \bibinfo {author} {\bibfnamefont {R.~J.}\ \bibnamefont {deBoer}}, \bibinfo
  {author} {\bibfnamefont {G.}~\bibnamefont {Imbriani}}, \bibinfo {author}
  {\bibfnamefont {A.}~\bibnamefont {Best}}, \bibinfo {author} {\bibfnamefont
  {A.}~\bibnamefont {Kontos}}, \bibinfo {author} {\bibfnamefont {P.~J.}\
  \bibnamefont {LeBlanc}}, \bibinfo {author} {\bibfnamefont {E.}~\bibnamefont
  {Uberseder}}, \ and\ \bibinfo {author} {\bibfnamefont {M.}~\bibnamefont
  {Wiescher}},\ }\href {\doibase 10.1103/PhysRevC.93.055806} {\bibfield
  {journal} {\bibinfo  {journal} {Phys. Rev. C}\ }\textbf {\bibinfo {volume}
  {93}},\ \bibinfo {pages} {055806} (\bibinfo {year} {2016})}\BibitemShut
  {NoStop}%
\bibitem [{\citenamefont {Frentz}\ \emph {et~al.}(2022)\citenamefont {Frentz},
  \citenamefont {Aprahamian}, \citenamefont {Boeltzig}, \citenamefont
  {Borgwardt}, \citenamefont {Clark}, \citenamefont {deBoer}, \citenamefont
  {Gilardy}, \citenamefont {G\"orres}, \citenamefont {Hanhardt}, \citenamefont
  {Henderson}, \citenamefont {Howard}, \citenamefont {Kadlecek}, \citenamefont
  {Liu}, \citenamefont {Macon}, \citenamefont {Moylan}, \citenamefont
  {Reingold}, \citenamefont {Robertson}, \citenamefont {Seymour}, \citenamefont
  {Strauss}, \citenamefont {Strieder}, \citenamefont {Vande~Kolk},\ and\
  \citenamefont {Wiescher}}]{Frentz202214N}%
  \BibitemOpen
  \bibfield  {author} {\bibinfo {author} {\bibfnamefont {B.}~\bibnamefont
  {Frentz}}, \bibinfo {author} {\bibfnamefont {A.}~\bibnamefont {Aprahamian}},
  \bibinfo {author} {\bibfnamefont {A.}~\bibnamefont {Boeltzig}}, \bibinfo
  {author} {\bibfnamefont {T.}~\bibnamefont {Borgwardt}}, \bibinfo {author}
  {\bibfnamefont {A.~M.}\ \bibnamefont {Clark}}, \bibinfo {author}
  {\bibfnamefont {R.~J.}\ \bibnamefont {deBoer}}, \bibinfo {author}
  {\bibfnamefont {G.}~\bibnamefont {Gilardy}}, \bibinfo {author} {\bibfnamefont
  {J.}~\bibnamefont {G\"orres}}, \bibinfo {author} {\bibfnamefont
  {M.}~\bibnamefont {Hanhardt}}, \bibinfo {author} {\bibfnamefont {S.~L.}\
  \bibnamefont {Henderson}}, \bibinfo {author} {\bibfnamefont {K.~B.}\
  \bibnamefont {Howard}}, \bibinfo {author} {\bibfnamefont {T.}~\bibnamefont
  {Kadlecek}}, \bibinfo {author} {\bibfnamefont {Q.}~\bibnamefont {Liu}},
  \bibinfo {author} {\bibfnamefont {K.~T.}\ \bibnamefont {Macon}}, \bibinfo
  {author} {\bibfnamefont {S.}~\bibnamefont {Moylan}}, \bibinfo {author}
  {\bibfnamefont {C.~S.}\ \bibnamefont {Reingold}}, \bibinfo {author}
  {\bibfnamefont {D.}~\bibnamefont {Robertson}}, \bibinfo {author}
  {\bibfnamefont {C.}~\bibnamefont {Seymour}}, \bibinfo {author} {\bibfnamefont
  {S.~Y.}\ \bibnamefont {Strauss}}, \bibinfo {author} {\bibfnamefont
  {F.}~\bibnamefont {Strieder}}, \bibinfo {author} {\bibfnamefont
  {B.}~\bibnamefont {Vande~Kolk}}, \ and\ \bibinfo {author} {\bibfnamefont
  {M.}~\bibnamefont {Wiescher}},\ }\href {\doibase 10.1103/PhysRevC.106.065803}
  {\bibfield  {journal} {\bibinfo  {journal} {Phys. Rev. C}\ }\textbf {\bibinfo
  {volume} {106}},\ \bibinfo {pages} {065803} (\bibinfo {year}
  {2022})}\BibitemShut {NoStop}%
\bibitem [{\citenamefont {Wagner}\ \emph {et~al.}(2018)\citenamefont {Wagner},
  \citenamefont {Akhmadaliev}, \citenamefont {Anders}, \citenamefont
  {Bemmerer}, \citenamefont {Caciolli}, \citenamefont {Gohl}, \citenamefont
  {Grieger}, \citenamefont {Junghans}, \citenamefont {Marta}, \citenamefont
  {Munnik}, \citenamefont {Reinhardt}, \citenamefont {Reinicke}, \citenamefont
  {R\"oder}, \citenamefont {Schmidt}, \citenamefont {Schwengner}, \citenamefont
  {Serfling}, \citenamefont {Tak\'acs}, \citenamefont {Sz\"ucs}, \citenamefont
  {Vomiero}, \citenamefont {Wagner},\ and\ \citenamefont
  {Zuber}}]{Wagner201814N}%
  \BibitemOpen
  \bibfield  {author} {\bibinfo {author} {\bibfnamefont {L.}~\bibnamefont
  {Wagner}}, \bibinfo {author} {\bibfnamefont {S.}~\bibnamefont {Akhmadaliev}},
  \bibinfo {author} {\bibfnamefont {M.}~\bibnamefont {Anders}}, \bibinfo
  {author} {\bibfnamefont {D.}~\bibnamefont {Bemmerer}}, \bibinfo {author}
  {\bibfnamefont {A.}~\bibnamefont {Caciolli}}, \bibinfo {author}
  {\bibfnamefont {S.}~\bibnamefont {Gohl}}, \bibinfo {author} {\bibfnamefont
  {M.}~\bibnamefont {Grieger}}, \bibinfo {author} {\bibfnamefont
  {A.}~\bibnamefont {Junghans}}, \bibinfo {author} {\bibfnamefont
  {M.}~\bibnamefont {Marta}}, \bibinfo {author} {\bibfnamefont
  {F.}~\bibnamefont {Munnik}}, \bibinfo {author} {\bibfnamefont {T.~P.}\
  \bibnamefont {Reinhardt}}, \bibinfo {author} {\bibfnamefont {S.}~\bibnamefont
  {Reinicke}}, \bibinfo {author} {\bibfnamefont {M.}~\bibnamefont {R\"oder}},
  \bibinfo {author} {\bibfnamefont {K.}~\bibnamefont {Schmidt}}, \bibinfo
  {author} {\bibfnamefont {R.}~\bibnamefont {Schwengner}}, \bibinfo {author}
  {\bibfnamefont {M.}~\bibnamefont {Serfling}}, \bibinfo {author}
  {\bibfnamefont {M.~P.}\ \bibnamefont {Tak\'acs}}, \bibinfo {author}
  {\bibfnamefont {T.}~\bibnamefont {Sz\"ucs}}, \bibinfo {author} {\bibfnamefont
  {A.}~\bibnamefont {Vomiero}}, \bibinfo {author} {\bibfnamefont
  {A.}~\bibnamefont {Wagner}}, \ and\ \bibinfo {author} {\bibfnamefont
  {K.}~\bibnamefont {Zuber}},\ }\href {\doibase 10.1103/PhysRevC.97.015801}
  {\bibfield  {journal} {\bibinfo  {journal} {Phys. Rev. C}\ }\textbf {\bibinfo
  {volume} {97}},\ \bibinfo {pages} {015801} (\bibinfo {year}
  {2018})}\BibitemShut {NoStop}%
\bibitem [{\citenamefont {Gy\"urky}\ \emph {et~al.}(2022)\citenamefont
  {Gy\"urky}, \citenamefont {Hal\'asz}, \citenamefont {Kiss}, \citenamefont
  {Sz\"ucs},\ and\ \citenamefont {F\"ul\"op}}]{ATOMKI14N}%
  \BibitemOpen
  \bibfield  {author} {\bibinfo {author} {\bibfnamefont {G.}~\bibnamefont
  {Gy\"urky}}, \bibinfo {author} {\bibfnamefont {Z.}~\bibnamefont {Hal\'asz}},
  \bibinfo {author} {\bibfnamefont {G.~G.}\ \bibnamefont {Kiss}}, \bibinfo
  {author} {\bibfnamefont {T.}~\bibnamefont {Sz\"ucs}}, \ and\ \bibinfo
  {author} {\bibfnamefont {Z.}~\bibnamefont {F\"ul\"op}},\ }\href {\doibase
  10.1103/PhysRevC.105.L022801} {\bibfield  {journal} {\bibinfo  {journal}
  {Phys. Rev. C}\ }\textbf {\bibinfo {volume} {105}},\ \bibinfo {pages}
  {L022801} (\bibinfo {year} {2022})}\BibitemShut {NoStop}%
\bibitem [{\citenamefont {Acharya}\ \emph {et~al.}(2024)\citenamefont
  {Acharya}, \citenamefont {Aliotta}, \citenamefont {Balantekin}, \citenamefont
  {Bemmerer}, \citenamefont {Bertulani}, \citenamefont {Best}, \citenamefont
  {Brune}, \citenamefont {Buompane}, \citenamefont {Cavanna}, \citenamefont
  {Chen}, \citenamefont {Colgan}, \citenamefont {Czarnecki}, \citenamefont
  {Davids}, \citenamefont {deBoer}, \citenamefont {Delahaye}, \citenamefont
  {Depalo}, \citenamefont {Garc¨ªa}, \citenamefont {Johnson}, \citenamefont
  {Gazit}, \citenamefont {Gialanella}, \citenamefont {Greife}, \citenamefont
  {Guffanti}, \citenamefont {Guglielmetti}, \citenamefont {Hambleton},
  \citenamefont {Haxton}, \citenamefont {Herrera}, \citenamefont {Huang},
  \citenamefont {Iliadis}, \citenamefont {Kravvaris}, \citenamefont {Cognata},
  \citenamefont {Langanke}, \citenamefont {Marcucci}, \citenamefont {Nagayama},
  \citenamefont {Nollett}, \citenamefont {Odell}, \citenamefont {Gann},
  \citenamefont {Piatti}, \citenamefont {Pinsonneault}, \citenamefont
  {Platter}, \citenamefont {Robertson}, \citenamefont {Rupak}, \citenamefont
  {Serenelli}, \citenamefont {Sferrazza}, \citenamefont {Sz¨¹cs}, \citenamefont
  {Tang}, \citenamefont {Tumino}, \citenamefont {Villante}, \citenamefont
  {Walker-Loud}, \citenamefont {Zhang},\ and\ \citenamefont {Zuber}}]{SF-III}%
  \BibitemOpen
  \bibfield  {author} {\bibinfo {author} {\bibfnamefont {B.}~\bibnamefont
  {Acharya}}, \bibinfo {author} {\bibfnamefont {M.}~\bibnamefont {Aliotta}},
  \bibinfo {author} {\bibfnamefont {A.~B.}\ \bibnamefont {Balantekin}},
  \bibinfo {author} {\bibfnamefont {D.}~\bibnamefont {Bemmerer}}, \bibinfo
  {author} {\bibfnamefont {C.~A.}\ \bibnamefont {Bertulani}}, \bibinfo {author}
  {\bibfnamefont {A.}~\bibnamefont {Best}}, \bibinfo {author} {\bibfnamefont
  {C.~R.}\ \bibnamefont {Brune}}, \bibinfo {author} {\bibfnamefont
  {R.}~\bibnamefont {Buompane}}, \bibinfo {author} {\bibfnamefont
  {F.}~\bibnamefont {Cavanna}}, \bibinfo {author} {\bibfnamefont {J.~W.}\
  \bibnamefont {Chen}}, \bibinfo {author} {\bibfnamefont {J.}~\bibnamefont
  {Colgan}}, \bibinfo {author} {\bibfnamefont {A.}~\bibnamefont {Czarnecki}},
  \bibinfo {author} {\bibfnamefont {B.}~\bibnamefont {Davids}}, \bibinfo
  {author} {\bibfnamefont {R.~J.}\ \bibnamefont {deBoer}}, \bibinfo {author}
  {\bibfnamefont {F.}~\bibnamefont {Delahaye}}, \bibinfo {author}
  {\bibfnamefont {R.}~\bibnamefont {Depalo}}, \bibinfo {author} {\bibfnamefont
  {A.}~\bibnamefont {Garc¨ªa}}, \bibinfo {author} {\bibfnamefont {M.~G.}\
  \bibnamefont {Johnson}}, \bibinfo {author} {\bibfnamefont {D.}~\bibnamefont
  {Gazit}}, \bibinfo {author} {\bibfnamefont {L.}~\bibnamefont {Gialanella}},
  \bibinfo {author} {\bibfnamefont {U.}~\bibnamefont {Greife}}, \bibinfo
  {author} {\bibfnamefont {D.}~\bibnamefont {Guffanti}}, \bibinfo {author}
  {\bibfnamefont {A.}~\bibnamefont {Guglielmetti}}, \bibinfo {author}
  {\bibfnamefont {K.}~\bibnamefont {Hambleton}}, \bibinfo {author}
  {\bibfnamefont {W.~C.}\ \bibnamefont {Haxton}}, \bibinfo {author}
  {\bibfnamefont {Y.}~\bibnamefont {Herrera}}, \bibinfo {author} {\bibfnamefont
  {M.}~\bibnamefont {Huang}}, \bibinfo {author} {\bibfnamefont
  {C.}~\bibnamefont {Iliadis}}, \bibinfo {author} {\bibfnamefont
  {K.}~\bibnamefont {Kravvaris}}, \bibinfo {author} {\bibfnamefont {M.~L.}\
  \bibnamefont {Cognata}}, \bibinfo {author} {\bibfnamefont {K.}~\bibnamefont
  {Langanke}}, \bibinfo {author} {\bibfnamefont {L.~E.}\ \bibnamefont
  {Marcucci}}, \bibinfo {author} {\bibfnamefont {T.}~\bibnamefont {Nagayama}},
  \bibinfo {author} {\bibfnamefont {K.~M.}\ \bibnamefont {Nollett}}, \bibinfo
  {author} {\bibfnamefont {D.}~\bibnamefont {Odell}}, \bibinfo {author}
  {\bibfnamefont {G.~D.~O.}\ \bibnamefont {Gann}}, \bibinfo {author}
  {\bibfnamefont {D.}~\bibnamefont {Piatti}}, \bibinfo {author} {\bibfnamefont
  {M.}~\bibnamefont {Pinsonneault}}, \bibinfo {author} {\bibfnamefont
  {L.}~\bibnamefont {Platter}}, \bibinfo {author} {\bibfnamefont {R.~G.~H.}\
  \bibnamefont {Robertson}}, \bibinfo {author} {\bibfnamefont {G.}~\bibnamefont
  {Rupak}}, \bibinfo {author} {\bibfnamefont {A.}~\bibnamefont {Serenelli}},
  \bibinfo {author} {\bibfnamefont {M.}~\bibnamefont {Sferrazza}}, \bibinfo
  {author} {\bibfnamefont {T.}~\bibnamefont {Sz¨¹cs}}, \bibinfo {author}
  {\bibfnamefont {X.}~\bibnamefont {Tang}}, \bibinfo {author} {\bibfnamefont
  {A.}~\bibnamefont {Tumino}}, \bibinfo {author} {\bibfnamefont {F.~L.}\
  \bibnamefont {Villante}}, \bibinfo {author} {\bibfnamefont {A.}~\bibnamefont
  {Walker-Loud}}, \bibinfo {author} {\bibfnamefont {X.}~\bibnamefont {Zhang}},
  \ and\ \bibinfo {author} {\bibfnamefont {K.}~\bibnamefont {Zuber}},\ }\href
  {https://arxiv.org/abs/2405.06470} {\enquote {\bibinfo {title} {Solar fusion
  iii: New data and theory for hydrogen-burning stars},}\ } (\bibinfo {year}
  {2024}),\ \Eprint {http://arxiv.org/abs/2405.06470} {arXiv:2405.06470
  [astro-ph.SR]} \BibitemShut {NoStop}%
\bibitem [{\citenamefont {Chen}\ \emph {et~al.}(2025)\citenamefont {Chen},
  \citenamefont {Su}, \citenamefont {Shen}, \citenamefont {Zhang},
  \citenamefont {He}, \citenamefont {Chen}, \citenamefont {Wang}, \citenamefont
  {Shen}, \citenamefont {Lin}, \citenamefont {Song}, \citenamefont {Zhang},
  \citenamefont {Wang}, \citenamefont {Jiang}, \citenamefont {Wang},
  \citenamefont {Huang}, \citenamefont {Qin}, \citenamefont {Liu},
  \citenamefont {Sheng}, \citenamefont {Chen}, \citenamefont {Lu},
  \citenamefont {Li}, \citenamefont {Dong}, \citenamefont {Jiang},
  \citenamefont {Zhang}, \citenamefont {Zhang}, \citenamefont {Tian},
  \citenamefont {Xiao}, \citenamefont {Zhang}, \citenamefont {Li},
  \citenamefont {Han}, \citenamefont {Wei}, \citenamefont {Li}, \citenamefont
  {An}, \citenamefont {Lin}, \citenamefont {Liao}, \citenamefont {Liu},
  \citenamefont {Zhang}, \citenamefont {Qiu}, \citenamefont {Xu}, \citenamefont
  {Jin}, \citenamefont {Lu}, \citenamefont {Chen}, \citenamefont {Nan},
  \citenamefont {Wang}, \citenamefont {Li}, \citenamefont {Guo}, \citenamefont
  {Gu},\ and\ \citenamefont {Liu}}]{chen202414npg}%
  \BibitemOpen
  \bibfield  {author} {\bibinfo {author} {\bibfnamefont {X.}~\bibnamefont
  {Chen}}, \bibinfo {author} {\bibfnamefont {J.}~\bibnamefont {Su}}, \bibinfo
  {author} {\bibfnamefont {Y.~P.}\ \bibnamefont {Shen}}, \bibinfo {author}
  {\bibfnamefont {L.~Y.}\ \bibnamefont {Zhang}}, \bibinfo {author}
  {\bibfnamefont {J.~J.}\ \bibnamefont {He}}, \bibinfo {author} {\bibfnamefont
  {S.~Z.}\ \bibnamefont {Chen}}, \bibinfo {author} {\bibfnamefont
  {S.}~\bibnamefont {Wang}}, \bibinfo {author} {\bibfnamefont {Z.~L.}\
  \bibnamefont {Shen}}, \bibinfo {author} {\bibfnamefont {S.}~\bibnamefont
  {Lin}}, \bibinfo {author} {\bibfnamefont {L.~Y.}\ \bibnamefont {Song}},
  \bibinfo {author} {\bibfnamefont {H.}~\bibnamefont {Zhang}}, \bibinfo
  {author} {\bibfnamefont {L.~H.}\ \bibnamefont {Wang}}, \bibinfo {author}
  {\bibfnamefont {X.~Z.}\ \bibnamefont {Jiang}}, \bibinfo {author}
  {\bibfnamefont {L.}~\bibnamefont {Wang}}, \bibinfo {author} {\bibfnamefont
  {Y.~T.}\ \bibnamefont {Huang}}, \bibinfo {author} {\bibfnamefont {Z.~W.}\
  \bibnamefont {Qin}}, \bibinfo {author} {\bibfnamefont {F.~C.}\ \bibnamefont
  {Liu}}, \bibinfo {author} {\bibfnamefont {Y.~D.}\ \bibnamefont {Sheng}},
  \bibinfo {author} {\bibfnamefont {Y.~J.}\ \bibnamefont {Chen}}, \bibinfo
  {author} {\bibfnamefont {Y.~L.}\ \bibnamefont {Lu}}, \bibinfo {author}
  {\bibfnamefont {X.~Y.}\ \bibnamefont {Li}}, \bibinfo {author} {\bibfnamefont
  {J.~Y.}\ \bibnamefont {Dong}}, \bibinfo {author} {\bibfnamefont {Y.~C.}\
  \bibnamefont {Jiang}}, \bibinfo {author} {\bibfnamefont {Y.~Q.}\ \bibnamefont
  {Zhang}}, \bibinfo {author} {\bibfnamefont {Y.}~\bibnamefont {Zhang}},
  \bibinfo {author} {\bibfnamefont {J.~W.}\ \bibnamefont {Tian}}, \bibinfo
  {author} {\bibfnamefont {D.}~\bibnamefont {Xiao}}, \bibinfo {author}
  {\bibfnamefont {Y.}~\bibnamefont {Zhang}}, \bibinfo {author} {\bibfnamefont
  {Z.~M.}\ \bibnamefont {Li}}, \bibinfo {author} {\bibfnamefont {X.~C.}\
  \bibnamefont {Han}}, \bibinfo {author} {\bibfnamefont {J.~J.}\ \bibnamefont
  {Wei}}, \bibinfo {author} {\bibfnamefont {H.}~\bibnamefont {Li}}, \bibinfo
  {author} {\bibfnamefont {Z.}~\bibnamefont {An}}, \bibinfo {author}
  {\bibfnamefont {W.~P.}\ \bibnamefont {Lin}}, \bibinfo {author} {\bibfnamefont
  {B.}~\bibnamefont {Liao}}, \bibinfo {author} {\bibfnamefont {H.~N.}\
  \bibnamefont {Liu}}, \bibinfo {author} {\bibfnamefont {F.~S.}\ \bibnamefont
  {Zhang}}, \bibinfo {author} {\bibfnamefont {M.~L.}\ \bibnamefont {Qiu}},
  \bibinfo {author} {\bibfnamefont {C.}~\bibnamefont {Xu}}, \bibinfo {author}
  {\bibfnamefont {S.~L.}\ \bibnamefont {Jin}}, \bibinfo {author} {\bibfnamefont
  {F.}~\bibnamefont {Lu}}, \bibinfo {author} {\bibfnamefont {J.~F.}\
  \bibnamefont {Chen}}, \bibinfo {author} {\bibfnamefont {W.}~\bibnamefont
  {Nan}}, \bibinfo {author} {\bibfnamefont {Y.~B.}\ \bibnamefont {Wang}},
  \bibinfo {author} {\bibfnamefont {Z.~H.}\ \bibnamefont {Li}}, \bibinfo
  {author} {\bibfnamefont {B.}~\bibnamefont {Guo}}, \bibinfo {author}
  {\bibfnamefont {Y.~Q.}\ \bibnamefont {Gu}}, \ and\ \bibinfo {author}
  {\bibfnamefont {W.~P.}\ \bibnamefont {Liu}},\ }\href {\doibase
  10.1103/q756-hzmt} {\bibfield  {journal} {\bibinfo  {journal} {Phys. Rev.
  Lett.}\ }\textbf {\bibinfo {volume} {135}},\ \bibinfo {pages} {232701}
  (\bibinfo {year} {2025})}\BibitemShut {NoStop}%
\bibitem [{\citenamefont {Angulo}\ and\ \citenamefont
  {Descouvemont}(2001)}]{ANGULO2001755}%
  \BibitemOpen
  \bibfield  {author} {\bibinfo {author} {\bibfnamefont {C.}~\bibnamefont
  {Angulo}}\ and\ \bibinfo {author} {\bibfnamefont {P.}~\bibnamefont
  {Descouvemont}},\ }\href {\doibase
  https://doi.org/10.1016/S0375-9474(00)00696-5} {\bibfield  {journal}
  {\bibinfo  {journal} {Nuclear Physics A}\ }\textbf {\bibinfo {volume}
  {690}},\ \bibinfo {pages} {755} (\bibinfo {year} {2001})}\BibitemShut
  {NoStop}%
\bibitem [{\citenamefont {Mukhamedzhanov}\ \emph {et~al.}(2003)\citenamefont
  {Mukhamedzhanov}, \citenamefont {B\'em}, \citenamefont {Brown}, \citenamefont
  {Burjan}, \citenamefont {Gagliardi}, \citenamefont {Kroha}, \citenamefont
  {Nov\'ak}, \citenamefont {Nunes}, \citenamefont {Pisko\ifmmode~\check{r}\else
  \v{r}\fi{}}, \citenamefont {Pirlepesov}, \citenamefont {\ifmmode
  \check{S}\else \v{S}\fi{}ime\ifmmode~\check{c}\else \v{c}\fi{}kov\'a},
  \citenamefont {Tribble},\ and\ \citenamefont {Vincour}}]{Mukhamedzhanov2003}%
  \BibitemOpen
  \bibfield  {author} {\bibinfo {author} {\bibfnamefont {A.~M.}\ \bibnamefont
  {Mukhamedzhanov}}, \bibinfo {author} {\bibfnamefont {P.}~\bibnamefont
  {B\'em}}, \bibinfo {author} {\bibfnamefont {B.~A.}\ \bibnamefont {Brown}},
  \bibinfo {author} {\bibfnamefont {V.}~\bibnamefont {Burjan}}, \bibinfo
  {author} {\bibfnamefont {C.~A.}\ \bibnamefont {Gagliardi}}, \bibinfo {author}
  {\bibfnamefont {V.}~\bibnamefont {Kroha}}, \bibinfo {author} {\bibfnamefont
  {J.}~\bibnamefont {Nov\'ak}}, \bibinfo {author} {\bibfnamefont {F.~M.}\
  \bibnamefont {Nunes}}, \bibinfo {author} {\bibfnamefont {i.~c.~v.}\
  \bibnamefont {Pisko\ifmmode~\check{r}\else \v{r}\fi{}}}, \bibinfo {author}
  {\bibfnamefont {F.}~\bibnamefont {Pirlepesov}}, \bibinfo {author}
  {\bibfnamefont {E.}~\bibnamefont {\ifmmode \check{S}\else
  \v{S}\fi{}ime\ifmmode~\check{c}\else \v{c}\fi{}kov\'a}}, \bibinfo {author}
  {\bibfnamefont {R.~E.}\ \bibnamefont {Tribble}}, \ and\ \bibinfo {author}
  {\bibfnamefont {J.}~\bibnamefont {Vincour}},\ }\href {\doibase
  10.1103/PhysRevC.67.065804} {\bibfield  {journal} {\bibinfo  {journal} {Phys.
  Rev. C}\ }\textbf {\bibinfo {volume} {67}},\ \bibinfo {pages} {065804}
  (\bibinfo {year} {2003})}\BibitemShut {NoStop}%
\bibitem [{\citenamefont {Azuma}\ \emph {et~al.}(2010)\citenamefont {Azuma},
  \citenamefont {Uberseder}, \citenamefont {Simpson}, \citenamefont {Brune},
  \citenamefont {Costantini}, \citenamefont {de~Boer}, \citenamefont
  {G\"orres}, \citenamefont {Heil}, \citenamefont {LeBlanc}, \citenamefont
  {Ugalde},\ and\ \citenamefont {Wiescher}}]{AZURE}%
  \BibitemOpen
  \bibfield  {author} {\bibinfo {author} {\bibfnamefont {R.~E.}\ \bibnamefont
  {Azuma}}, \bibinfo {author} {\bibfnamefont {E.}~\bibnamefont {Uberseder}},
  \bibinfo {author} {\bibfnamefont {E.~C.}\ \bibnamefont {Simpson}}, \bibinfo
  {author} {\bibfnamefont {C.~R.}\ \bibnamefont {Brune}}, \bibinfo {author}
  {\bibfnamefont {H.}~\bibnamefont {Costantini}}, \bibinfo {author}
  {\bibfnamefont {R.~J.}\ \bibnamefont {de~Boer}}, \bibinfo {author}
  {\bibfnamefont {J.}~\bibnamefont {G\"orres}}, \bibinfo {author}
  {\bibfnamefont {M.}~\bibnamefont {Heil}}, \bibinfo {author} {\bibfnamefont
  {P.~J.}\ \bibnamefont {LeBlanc}}, \bibinfo {author} {\bibfnamefont
  {C.}~\bibnamefont {Ugalde}}, \ and\ \bibinfo {author} {\bibfnamefont
  {M.}~\bibnamefont {Wiescher}},\ }\href {\doibase 10.1103/PhysRevC.81.045805}
  {\bibfield  {journal} {\bibinfo  {journal} {Phys. Rev. C}\ }\textbf {\bibinfo
  {volume} {81}},\ \bibinfo {pages} {045805} (\bibinfo {year}
  {2010})}\BibitemShut {NoStop}%
\bibitem [{\citenamefont {Frentz}\ \emph {et~al.}(2021)\citenamefont {Frentz},
  \citenamefont {Aprahamian}, \citenamefont {Clark}, \citenamefont {deBoer},
  \citenamefont {Dulal}, \citenamefont {Enright}, \citenamefont {G\"orres},
  \citenamefont {Henderson}, \citenamefont {Hinnefeld}, \citenamefont {Howard},
  \citenamefont {Kelmar}, \citenamefont {Lee}, \citenamefont {Morales},
  \citenamefont {Moylan}, \citenamefont {Rahman}, \citenamefont {Tan},
  \citenamefont {Weghorn},\ and\ \citenamefont
  {Wiescher}}]{Frentz2021lifetime}%
  \BibitemOpen
  \bibfield  {author} {\bibinfo {author} {\bibfnamefont {B.}~\bibnamefont
  {Frentz}}, \bibinfo {author} {\bibfnamefont {A.}~\bibnamefont {Aprahamian}},
  \bibinfo {author} {\bibfnamefont {A.~M.}\ \bibnamefont {Clark}}, \bibinfo
  {author} {\bibfnamefont {R.~J.}\ \bibnamefont {deBoer}}, \bibinfo {author}
  {\bibfnamefont {C.}~\bibnamefont {Dulal}}, \bibinfo {author} {\bibfnamefont
  {J.~D.}\ \bibnamefont {Enright}}, \bibinfo {author} {\bibfnamefont
  {J.}~\bibnamefont {G\"orres}}, \bibinfo {author} {\bibfnamefont {S.~L.}\
  \bibnamefont {Henderson}}, \bibinfo {author} {\bibfnamefont {J.~D.}\
  \bibnamefont {Hinnefeld}}, \bibinfo {author} {\bibfnamefont {K.~B.}\
  \bibnamefont {Howard}}, \bibinfo {author} {\bibfnamefont {R.}~\bibnamefont
  {Kelmar}}, \bibinfo {author} {\bibfnamefont {K.}~\bibnamefont {Lee}},
  \bibinfo {author} {\bibfnamefont {L.}~\bibnamefont {Morales}}, \bibinfo
  {author} {\bibfnamefont {S.}~\bibnamefont {Moylan}}, \bibinfo {author}
  {\bibfnamefont {Z.}~\bibnamefont {Rahman}}, \bibinfo {author} {\bibfnamefont
  {W.}~\bibnamefont {Tan}}, \bibinfo {author} {\bibfnamefont {L.~E.}\
  \bibnamefont {Weghorn}}, \ and\ \bibinfo {author} {\bibfnamefont
  {M.}~\bibnamefont {Wiescher}},\ }\href {\doibase 10.1103/PhysRevC.103.045802}
  {\bibfield  {journal} {\bibinfo  {journal} {Phys. Rev. C}\ }\textbf {\bibinfo
  {volume} {103}},\ \bibinfo {pages} {045802} (\bibinfo {year}
  {2021})}\BibitemShut {NoStop}%
\bibitem [{\citenamefont {Dubovichenko}\ \emph {et~al.}(2020)\citenamefont
  {Dubovichenko}, \citenamefont {Burkova}, \citenamefont
  {Dzhazairov-Kakhramanov},\ and\ \citenamefont {Beysenov}}]{Dubovichenko14N}%
  \BibitemOpen
  \bibfield  {author} {\bibinfo {author} {\bibfnamefont {S.}~\bibnamefont
  {Dubovichenko}}, \bibinfo {author} {\bibfnamefont {N.}~\bibnamefont
  {Burkova}}, \bibinfo {author} {\bibfnamefont {A.}~\bibnamefont
  {Dzhazairov-Kakhramanov}}, \ and\ \bibinfo {author} {\bibfnamefont
  {B.}~\bibnamefont {Beysenov}},\ }\href {\doibase 10.1142/S0218301319300078}
  {\bibfield  {journal} {\bibinfo  {journal} {International Journal of Modern
  Physics E}\ }\textbf {\bibinfo {volume} {29}},\ \bibinfo {pages} {1930007}
  (\bibinfo {year} {2020})},\ \Eprint
  {http://arxiv.org/abs/https://doi.org/10.1142/S0218301319300078}
  {https://doi.org/10.1142/S0218301319300078} \BibitemShut {NoStop}%
\bibitem [{GSM()}]{GSMbook}%
  \BibitemOpen
  \href@noop {} {}\bibinfo {note} {N. Michel and M. P{\l}oszajczak, {\it Gamow
  Shell Model - The Unified Theory of Nuclear Structure and Reactions}, Lecture
  Notes in Physics Vol. 983 (Springer, Cham, 2021)}\BibitemShut {NoStop}%
\bibitem [{\citenamefont {Jaganathen}\ \emph {et~al.}(2014)\citenamefont
  {Jaganathen}, \citenamefont {Michel},\ and\ \citenamefont
  {P{\l}oszajczak}}]{Jaganathen14}%
  \BibitemOpen
  \bibfield  {author} {\bibinfo {author} {\bibfnamefont {Y.}~\bibnamefont
  {Jaganathen}}, \bibinfo {author} {\bibfnamefont {N.}~\bibnamefont {Michel}},
  \ and\ \bibinfo {author} {\bibfnamefont {M.}~\bibnamefont {P{\l}oszajczak}},\
  }\href@noop {} {\bibfield  {journal} {\bibinfo  {journal} {Phys. Rev. C}\
  }\textbf {\bibinfo {volume} {89}},\ \bibinfo {pages} {034624} (\bibinfo
  {year} {2014})}\BibitemShut {NoStop}%
\bibitem [{\citenamefont {Mercenne}\ \emph {et~al.}(2019)\citenamefont
  {Mercenne}, \citenamefont {Michel},\ and\ \citenamefont
  {P{\l}oszajczak}}]{Mercenne19}%
  \BibitemOpen
  \bibfield  {author} {\bibinfo {author} {\bibfnamefont {A.}~\bibnamefont
  {Mercenne}}, \bibinfo {author} {\bibfnamefont {N.}~\bibnamefont {Michel}}, \
  and\ \bibinfo {author} {\bibfnamefont {M.}~\bibnamefont {P{\l}oszajczak}},\
  }\href@noop {} {\bibfield  {journal} {\bibinfo  {journal} {Phys. Rev. C}\
  }\textbf {\bibinfo {volume} {99}},\ \bibinfo {pages} {044606} (\bibinfo
  {year} {2019})}\BibitemShut {NoStop}%
\bibitem [{\citenamefont {Linares~Fernandez}\ \emph {et~al.}(2023)\citenamefont
  {Linares~Fernandez}, \citenamefont {Michel}, \citenamefont {P\l{}oszajczak},\
  and\ \citenamefont {Mercenne}}]{Fernandez23}%
  \BibitemOpen
  \bibfield  {author} {\bibinfo {author} {\bibfnamefont {J.~P.}\ \bibnamefont
  {Linares~Fernandez}}, \bibinfo {author} {\bibfnamefont {N.}~\bibnamefont
  {Michel}}, \bibinfo {author} {\bibfnamefont {M.}~\bibnamefont
  {P\l{}oszajczak}}, \ and\ \bibinfo {author} {\bibfnamefont {A.}~\bibnamefont
  {Mercenne}},\ }\href {\doibase 10.1103/PhysRevC.108.044616} {\bibfield
  {journal} {\bibinfo  {journal} {Phys. Rev. C}\ }\textbf {\bibinfo {volume}
  {108}},\ \bibinfo {pages} {044616} (\bibinfo {year} {2023})}\BibitemShut
  {NoStop}%
\bibitem [{\citenamefont {Fossez}\ \emph {et~al.}(2015)\citenamefont {Fossez},
  \citenamefont {Michel}, \citenamefont {P\l{}oszajczak}, \citenamefont
  {Jaganathen},\ and\ \citenamefont {Id~Betan}}]{Fossez15}%
  \BibitemOpen
  \bibfield  {author} {\bibinfo {author} {\bibfnamefont {K.}~\bibnamefont
  {Fossez}}, \bibinfo {author} {\bibfnamefont {N.}~\bibnamefont {Michel}},
  \bibinfo {author} {\bibfnamefont {M.}~\bibnamefont {P\l{}oszajczak}},
  \bibinfo {author} {\bibfnamefont {Y.}~\bibnamefont {Jaganathen}}, \ and\
  \bibinfo {author} {\bibfnamefont {R.~M.}\ \bibnamefont {Id~Betan}},\ }\href
  {\doibase 10.1103/PhysRevC.91.034609} {\bibfield  {journal} {\bibinfo
  {journal} {Phys. Rev. C}\ }\textbf {\bibinfo {volume} {91}},\ \bibinfo
  {pages} {034609} (\bibinfo {year} {2015})}\BibitemShut {NoStop}%
\bibitem [{\citenamefont {Dong}\ \emph {et~al.}(2017)\citenamefont {Dong},
  \citenamefont {Michel}, \citenamefont {Fossez}, \citenamefont
  {P{\l}oszajczak}, \citenamefont {Jaganathen},\ and\ \citenamefont
  {Id~Betan}}]{dong17}%
  \BibitemOpen
  \bibfield  {author} {\bibinfo {author} {\bibfnamefont {G.~X.}\ \bibnamefont
  {Dong}}, \bibinfo {author} {\bibfnamefont {N.}~\bibnamefont {Michel}},
  \bibinfo {author} {\bibfnamefont {K.}~\bibnamefont {Fossez}}, \bibinfo
  {author} {\bibfnamefont {M.}~\bibnamefont {P{\l}oszajczak}}, \bibinfo
  {author} {\bibfnamefont {Y.}~\bibnamefont {Jaganathen}}, \ and\ \bibinfo
  {author} {\bibfnamefont {R.~M.}\ \bibnamefont {Id~Betan}},\ }\href {\doibase
  10.1088/1361-6471/aa5f24} {\bibfield  {journal} {\bibinfo  {journal} {J.
  Phys. G: Nucl. Part. Phys.}\ }\textbf {\bibinfo {volume} {44}},\ \bibinfo
  {pages} {045201} (\bibinfo {year} {2017})}\BibitemShut {NoStop}%
\bibitem [{\citenamefont {Dong}\ \emph {et~al.}(2022)\citenamefont {Dong},
  \citenamefont {Wang}, \citenamefont {Michel},\ and\ \citenamefont
  {P{\l}oszajczak}}]{dong22}%
  \BibitemOpen
  \bibfield  {author} {\bibinfo {author} {\bibfnamefont {G.~X.}\ \bibnamefont
  {Dong}}, \bibinfo {author} {\bibfnamefont {X.~B.}\ \bibnamefont {Wang}},
  \bibinfo {author} {\bibfnamefont {N.}~\bibnamefont {Michel}}, \ and\ \bibinfo
  {author} {\bibfnamefont {M.}~\bibnamefont {P{\l}oszajczak}},\ }\href@noop {}
  {\bibfield  {journal} {\bibinfo  {journal} {Phys. Pev. C}\ }\textbf {\bibinfo
  {volume} {105}},\ \bibinfo {pages} {064608} (\bibinfo {year}
  {2022})}\BibitemShut {NoStop}%
\bibitem [{\citenamefont {Dong}\ \emph {et~al.}(2023)\citenamefont {Dong},
  \citenamefont {Wang}, \citenamefont {Michel},\ and\ \citenamefont
  {P{\l}oszajczak}}]{dong23}%
  \BibitemOpen
  \bibfield  {author} {\bibinfo {author} {\bibfnamefont {G.~X.}\ \bibnamefont
  {Dong}}, \bibinfo {author} {\bibfnamefont {X.~B.}\ \bibnamefont {Wang}},
  \bibinfo {author} {\bibfnamefont {N.}~\bibnamefont {Michel}}, \ and\ \bibinfo
  {author} {\bibfnamefont {M.}~\bibnamefont {P{\l}oszajczak}},\ }\href@noop {}
  {\bibfield  {journal} {\bibinfo  {journal} {Phys. Rev. C}\ }\textbf {\bibinfo
  {volume} {107}},\ \bibinfo {pages} {044613} (\bibinfo {year}
  {2023})}\BibitemShut {NoStop}%
\bibitem [{\citenamefont {Dong}\ \emph {et~al.}(2024)\citenamefont {Dong},
  \citenamefont {Wang}, \citenamefont {Michel},\ and\ \citenamefont
  {P\l{}oszajczak}}]{DongGX2024}%
  \BibitemOpen
  \bibfield  {author} {\bibinfo {author} {\bibfnamefont {G.~X.}\ \bibnamefont
  {Dong}}, \bibinfo {author} {\bibfnamefont {X.~B.}\ \bibnamefont {Wang}},
  \bibinfo {author} {\bibfnamefont {N.}~\bibnamefont {Michel}}, \ and\ \bibinfo
  {author} {\bibfnamefont {M.}~\bibnamefont {P\l{}oszajczak}},\ }\href
  {\doibase 10.1103/PhysRevC.110.L061601} {\bibfield  {journal} {\bibinfo
  {journal} {Phys. Rev. C}\ }\textbf {\bibinfo {volume} {110}},\ \bibinfo
  {pages} {L061601} (\bibinfo {year} {2024})}\BibitemShut {NoStop}%
\bibitem [{sup()}]{supplement}%
  \BibitemOpen
  \href@noop {} {}\bibinfo {note} {See Supplemental Material at [URL inserted
  by publisher] for the formulations of GSM-CC in more details; the one-body
  potential and the two-body interaction used for GSM-CC
  Hamiltonian}\BibitemShut {NoStop}%
\bibitem [{\citenamefont {Michel}\ \emph {et~al.}(2009)\citenamefont {Michel},
  \citenamefont {Nazarewicz}, \citenamefont {P{\l}oszajczak},\ and\
  \citenamefont {Vertse}}]{Michel09}%
  \BibitemOpen
  \bibfield  {author} {\bibinfo {author} {\bibfnamefont {N.}~\bibnamefont
  {Michel}}, \bibinfo {author} {\bibfnamefont {W.}~\bibnamefont {Nazarewicz}},
  \bibinfo {author} {\bibfnamefont {M.}~\bibnamefont {P{\l}oszajczak}}, \ and\
  \bibinfo {author} {\bibfnamefont {T.}~\bibnamefont {Vertse}},\ }\href@noop {}
  {\bibfield  {journal} {\bibinfo  {journal} {J. Phys. G: Nucl. Part. Phys.}\
  }\textbf {\bibinfo {volume} {36}},\ \bibinfo {pages} {013101} (\bibinfo
  {year} {2009})}\BibitemShut {NoStop}%
\bibitem [{\citenamefont {Suzuki}\ and\ \citenamefont {Ikeda}(1988)}]{Ikeda88}%
  \BibitemOpen
  \bibfield  {author} {\bibinfo {author} {\bibfnamefont {Y.}~\bibnamefont
  {Suzuki}}\ and\ \bibinfo {author} {\bibfnamefont {K.}~\bibnamefont {Ikeda}},\
  }\href@noop {} {\bibfield  {journal} {\bibinfo  {journal} {Phys. Pev. C}\
  }\textbf {\bibinfo {volume} {38}},\ \bibinfo {pages} {410} (\bibinfo {year}
  {1988})}\BibitemShut {NoStop}%
\bibitem [{\citenamefont {Berggren}(1968)}]{rf:4}%
  \BibitemOpen
  \bibfield  {author} {\bibinfo {author} {\bibfnamefont {T.}~\bibnamefont
  {Berggren}},\ }\href@noop {} {\bibfield  {journal} {\bibinfo  {journal}
  {Nucl. Phys. A}\ }\textbf {\bibinfo {volume} {109}},\ \bibinfo {pages} {265}
  (\bibinfo {year} {1968})}\BibitemShut {NoStop}%
\bibitem [{Mic()}]{Michel02RIB}%
  \BibitemOpen
  \href@noop {} {}\bibinfo {note} {N. Michel, W. Nazarewicz, M. P{\l}oszajczak,
  and K. Bennaceur, Phys. Rev. Lett. \textbf{89}, 042502 (2002). \\R. Id Betan,
  R. J. Liotta, N. Sandulescu, and T. Vertse, Phys. Rev. Lett. \textbf{89},
  042501 (2002)}\BibitemShut {NoStop}%
\bibitem [{\citenamefont {Michel}\ \emph {et~al.}(2003)\citenamefont {Michel},
  \citenamefont {Nazarewicz}, \citenamefont {P{\l}oszajczak},\ and\
  \citenamefont {Oko{\l}owicz}}]{Michel03}%
  \BibitemOpen
  \bibfield  {author} {\bibinfo {author} {\bibfnamefont {N.}~\bibnamefont
  {Michel}}, \bibinfo {author} {\bibfnamefont {W.}~\bibnamefont {Nazarewicz}},
  \bibinfo {author} {\bibfnamefont {M.}~\bibnamefont {P{\l}oszajczak}}, \ and\
  \bibinfo {author} {\bibfnamefont {J.}~\bibnamefont {Oko{\l}owicz}},\
  }\href@noop {} {\bibfield  {journal} {\bibinfo  {journal} {Phys. Rev. C}\
  }\textbf {\bibinfo {volume} {67}},\ \bibinfo {pages} {054311} (\bibinfo
  {year} {2003})}\BibitemShut {NoStop}%
\bibitem [{\citenamefont {Furutani}\ \emph {et~al.}(1978)\citenamefont
  {Furutani}, \citenamefont {Horiuchi},\ and\ \citenamefont
  {Tamagaki}}]{Furutani78}%
  \BibitemOpen
  \bibfield  {author} {\bibinfo {author} {\bibfnamefont {H.}~\bibnamefont
  {Furutani}}, \bibinfo {author} {\bibfnamefont {H.}~\bibnamefont {Horiuchi}},
  \ and\ \bibinfo {author} {\bibfnamefont {R.}~\bibnamefont {Tamagaki}},\
  }\href@noop {} {\bibfield  {journal} {\bibinfo  {journal} {Prog. Theor.
  Phys.}\ }\textbf {\bibinfo {volume} {60}},\ \bibinfo {pages} {307} (\bibinfo
  {year} {1978})}\BibitemShut {NoStop}%
\bibitem [{\citenamefont {Furutani}\ \emph {et~al.}(1979)\citenamefont
  {Furutani}, \citenamefont {Horiuchi},\ and\ \citenamefont
  {Tamagaki}}]{Furutani79}%
  \BibitemOpen
  \bibfield  {author} {\bibinfo {author} {\bibfnamefont {H.}~\bibnamefont
  {Furutani}}, \bibinfo {author} {\bibfnamefont {H.}~\bibnamefont {Horiuchi}},
  \ and\ \bibinfo {author} {\bibfnamefont {R.}~\bibnamefont {Tamagaki}},\
  }\href@noop {} {\bibfield  {journal} {\bibinfo  {journal} {Prog. Theor.
  Phys.}\ }\textbf {\bibinfo {volume} {62}},\ \bibinfo {pages} {981} (\bibinfo
  {year} {1979})}\BibitemShut {NoStop}%
\bibitem [{\citenamefont {Jaganathen}\ \emph {et~al.}(2017)\citenamefont
  {Jaganathen}, \citenamefont {Id~Betan}, \citenamefont {Michel}, \citenamefont
  {Nazarewicz},\ and\ \citenamefont {P\l{}oszajczak}}]{jaganathen_2017}%
  \BibitemOpen
  \bibfield  {author} {\bibinfo {author} {\bibfnamefont {Y.}~\bibnamefont
  {Jaganathen}}, \bibinfo {author} {\bibfnamefont {R.~M.}\ \bibnamefont
  {Id~Betan}}, \bibinfo {author} {\bibfnamefont {N.}~\bibnamefont {Michel}},
  \bibinfo {author} {\bibfnamefont {W.}~\bibnamefont {Nazarewicz}}, \ and\
  \bibinfo {author} {\bibfnamefont {M.}~\bibnamefont {P\l{}oszajczak}},\ }\href
  {\doibase 10.1103/PhysRevC.96.054316} {\bibfield  {journal} {\bibinfo
  {journal} {Phys. Rev. C}\ }\textbf {\bibinfo {volume} {96}},\ \bibinfo
  {pages} {054316} (\bibinfo {year} {2017})}\BibitemShut {NoStop}%
\bibitem [{\citenamefont {Xu}\ \emph {et~al.}(1994)\citenamefont {Xu},
  \citenamefont {Gagliardi}, \citenamefont {Tribble}, \citenamefont
  {Mukhamedzhanov},\ and\ \citenamefont {Timofeyuk}}]{XuHM1994}%
  \BibitemOpen
  \bibfield  {author} {\bibinfo {author} {\bibfnamefont {H.~M.}\ \bibnamefont
  {Xu}}, \bibinfo {author} {\bibfnamefont {C.~A.}\ \bibnamefont {Gagliardi}},
  \bibinfo {author} {\bibfnamefont {R.~E.}\ \bibnamefont {Tribble}}, \bibinfo
  {author} {\bibfnamefont {A.~M.}\ \bibnamefont {Mukhamedzhanov}}, \ and\
  \bibinfo {author} {\bibfnamefont {N.~K.}\ \bibnamefont {Timofeyuk}},\ }\href
  {\doibase 10.1103/PhysRevLett.73.2027} {\bibfield  {journal} {\bibinfo
  {journal} {Phys. Rev. Lett.}\ }\textbf {\bibinfo {volume} {73}},\ \bibinfo
  {pages} {2027} (\bibinfo {year} {1994})}\BibitemShut {NoStop}%
\bibitem [{\citenamefont {Vaclav}\ \emph {et~al.}(2020)\citenamefont {Vaclav},
  \citenamefont {Jaromir},\ and\ \citenamefont {Giuseppe}}]{Burjan20}%
  \BibitemOpen
  \bibfield  {author} {\bibinfo {author} {\bibfnamefont {B.}~\bibnamefont
  {Vaclav}}, \bibinfo {author} {\bibfnamefont {M.}~\bibnamefont {Jaromir}}, \
  and\ \bibinfo {author} {\bibfnamefont {D.}~\bibnamefont {Giuseppe}},\ }\href
  {\doibase 10.3389/fspas.2020.562466} {\bibfield  {journal} {\bibinfo
  {journal} {Front. Astron. Space Sci.}\ }\textbf {\bibinfo {volume} {7}},\
  \bibinfo {pages} {562466} (\bibinfo {year} {2020})}\BibitemShut {NoStop}%
\bibitem [{\citenamefont {Bertone}\ \emph {et~al.}(2002)\citenamefont
  {Bertone}, \citenamefont {Champagne}, \citenamefont {Boswell}, \citenamefont
  {Iliadis}, \citenamefont {Hale}, \citenamefont {Hansper},\ and\ \citenamefont
  {Powell}}]{Bertone2002}%
  \BibitemOpen
  \bibfield  {author} {\bibinfo {author} {\bibfnamefont {P.~F.}\ \bibnamefont
  {Bertone}}, \bibinfo {author} {\bibfnamefont {A.~E.}\ \bibnamefont
  {Champagne}}, \bibinfo {author} {\bibfnamefont {M.}~\bibnamefont {Boswell}},
  \bibinfo {author} {\bibfnamefont {C.}~\bibnamefont {Iliadis}}, \bibinfo
  {author} {\bibfnamefont {S.~E.}\ \bibnamefont {Hale}}, \bibinfo {author}
  {\bibfnamefont {V.~Y.}\ \bibnamefont {Hansper}}, \ and\ \bibinfo {author}
  {\bibfnamefont {D.~C.}\ \bibnamefont {Powell}},\ }\href {\doibase
  10.1103/PhysRevC.66.055804} {\bibfield  {journal} {\bibinfo  {journal} {Phys.
  Rev. C}\ }\textbf {\bibinfo {volume} {66}},\ \bibinfo {pages} {055804}
  (\bibinfo {year} {2002})}\BibitemShut {NoStop}%
\bibitem [{nnd()}]{nndc}%
  \BibitemOpen
  \href@noop {} {}\bibinfo {note} {NuDat 3.0 database (National Nuclear Data
  Center, Brookhaven National Laboratory),
  \url{http://www.nndc.bnl.gov/nudat3/}}\BibitemShut {NoStop}%
\bibitem [{\citenamefont {Stone}(2014)}]{stone2014table}%
  \BibitemOpen
  \bibfield  {author} {\bibinfo {author} {\bibfnamefont {N.}~\bibnamefont
  {Stone}},\ }\href@noop {} {\bibfield  {journal} {\bibinfo  {journal}
  {International Nuclear Data Committee Report INDC (NDS)-0658, Vienna,
  Austria}\ } (\bibinfo {year} {2014})}\BibitemShut {NoStop}%
\bibitem [{\citenamefont {Hornyak}(1975)}]{Hornyak75}%
  \BibitemOpen
  \bibfield  {author} {\bibinfo {author} {\bibfnamefont {W.~F.}\ \bibnamefont
  {Hornyak}},\ }\href@noop {} {\emph {\bibinfo {title} {Nuclear Structure}}}\
  (\bibinfo  {publisher} {Academic Press, New York},\ \bibinfo {year}
  {1975})\BibitemShut {NoStop}%
\bibitem [{\citenamefont {Ho}\ and\ \citenamefont {Coceva}(1988)}]{YKHo88}%
  \BibitemOpen
  \bibfield  {author} {\bibinfo {author} {\bibfnamefont {Y.~K.}\ \bibnamefont
  {Ho}}\ and\ \bibinfo {author} {\bibfnamefont {C.}~\bibnamefont {Coceva}},\
  }\href@noop {} {\bibfield  {journal} {\bibinfo  {journal} {J. Phys. G: Nucl.
  Phys.}\ }\textbf {\bibinfo {volume} {14}},\ \bibinfo {pages} {S207} (\bibinfo
  {year} {1988})}\BibitemShut {NoStop}%
\bibitem [{\citenamefont {Basilico}\ \emph {et~al.}(2023)\citenamefont
  {Basilico}, \citenamefont {Bellini}, \citenamefont {Benziger}, \citenamefont
  {Biondi}, \citenamefont {Caccianiga}, \citenamefont {Calaprice},
  \citenamefont {Caminata}, \citenamefont {Chepurnov}, \citenamefont
  {D'Angelo}, \citenamefont {Derbin}, \citenamefont {Di~Giacinto},
  \citenamefont {Di~Marcello}, \citenamefont {Ding}, \citenamefont
  {Di~Ludovico}, \citenamefont {Di~Noto}, \citenamefont {Drachnev},
  \citenamefont {Franco}, \citenamefont {Galbiati}, \citenamefont {Ghiano},
  \citenamefont {Giammarchi}, \citenamefont {Goretti}, \citenamefont {Gromov},
  \citenamefont {Guffanti}, \citenamefont {Ianni}, \citenamefont {Ianni},
  \citenamefont {Jany}, \citenamefont {Kobychev}, \citenamefont {Korga},
  \citenamefont {Kumaran}, \citenamefont {Laubenstein}, \citenamefont
  {Litvinovich}, \citenamefont {Lombardi}, \citenamefont {Lomskaya},
  \citenamefont {Ludhova}, \citenamefont {Machulin}, \citenamefont {Martyn},
  \citenamefont {Meroni}, \citenamefont {Miramonti}, \citenamefont {Misiaszek},
  \citenamefont {Muratova}, \citenamefont {Nugmanov}, \citenamefont {Oberauer},
  \citenamefont {Orekhov}, \citenamefont {Ortica}, \citenamefont {Pallavicini},
  \citenamefont {Pelicci}, \citenamefont {Penek}, \citenamefont {Pietrofaccia},
  \citenamefont {Pilipenko}, \citenamefont {Pocar}, \citenamefont {Raikov},
  \citenamefont {Ranalli}, \citenamefont {Ranucci}, \citenamefont {Razeto},
  \citenamefont {Re}, \citenamefont {Rossi}, \citenamefont {Sch\"onert},
  \citenamefont {Semenov}, \citenamefont {Settanta}, \citenamefont
  {Skorokhvatov}, \citenamefont {Singhal}, \citenamefont {Smirnov},
  \citenamefont {Sotnikov}, \citenamefont {Tartaglia}, \citenamefont {Testera},
  \citenamefont {Unzhakov}, \citenamefont {Villante}, \citenamefont {Vishneva},
  \citenamefont {Vogelaar}, \citenamefont {von Feilitzsch}, \citenamefont
  {Wojcik}, \citenamefont {Wurm}, \citenamefont {Zavatarelli}, \citenamefont
  {Zuber},\ and\ \citenamefont {Zuzel}}]{Basilico23}%
  \BibitemOpen
  \bibfield  {author} {\bibinfo {author} {\bibfnamefont {D.}~\bibnamefont
  {Basilico}}, \bibinfo {author} {\bibfnamefont {G.}~\bibnamefont {Bellini}},
  \bibinfo {author} {\bibfnamefont {J.}~\bibnamefont {Benziger}}, \bibinfo
  {author} {\bibfnamefont {R.}~\bibnamefont {Biondi}}, \bibinfo {author}
  {\bibfnamefont {B.}~\bibnamefont {Caccianiga}}, \bibinfo {author}
  {\bibfnamefont {F.}~\bibnamefont {Calaprice}}, \bibinfo {author}
  {\bibfnamefont {A.}~\bibnamefont {Caminata}}, \bibinfo {author}
  {\bibfnamefont {A.}~\bibnamefont {Chepurnov}}, \bibinfo {author}
  {\bibfnamefont {D.}~\bibnamefont {D'Angelo}}, \bibinfo {author}
  {\bibfnamefont {A.}~\bibnamefont {Derbin}}, \bibinfo {author} {\bibfnamefont
  {A.}~\bibnamefont {Di~Giacinto}}, \bibinfo {author} {\bibfnamefont
  {V.}~\bibnamefont {Di~Marcello}}, \bibinfo {author} {\bibfnamefont {X.~F.}\
  \bibnamefont {Ding}}, \bibinfo {author} {\bibfnamefont {A.}~\bibnamefont
  {Di~Ludovico}}, \bibinfo {author} {\bibfnamefont {L.}~\bibnamefont
  {Di~Noto}}, \bibinfo {author} {\bibfnamefont {I.}~\bibnamefont {Drachnev}},
  \bibinfo {author} {\bibfnamefont {D.}~\bibnamefont {Franco}}, \bibinfo
  {author} {\bibfnamefont {C.}~\bibnamefont {Galbiati}}, \bibinfo {author}
  {\bibfnamefont {C.}~\bibnamefont {Ghiano}}, \bibinfo {author} {\bibfnamefont
  {M.}~\bibnamefont {Giammarchi}}, \bibinfo {author} {\bibfnamefont
  {A.}~\bibnamefont {Goretti}}, \bibinfo {author} {\bibfnamefont
  {M.}~\bibnamefont {Gromov}}, \bibinfo {author} {\bibfnamefont
  {D.}~\bibnamefont {Guffanti}}, \bibinfo {author} {\bibfnamefont
  {A.}~\bibnamefont {Ianni}}, \bibinfo {author} {\bibfnamefont
  {A.}~\bibnamefont {Ianni}}, \bibinfo {author} {\bibfnamefont
  {A.}~\bibnamefont {Jany}}, \bibinfo {author} {\bibfnamefont {V.}~\bibnamefont
  {Kobychev}}, \bibinfo {author} {\bibfnamefont {G.}~\bibnamefont {Korga}},
  \bibinfo {author} {\bibfnamefont {S.}~\bibnamefont {Kumaran}}, \bibinfo
  {author} {\bibfnamefont {M.}~\bibnamefont {Laubenstein}}, \bibinfo {author}
  {\bibfnamefont {E.}~\bibnamefont {Litvinovich}}, \bibinfo {author}
  {\bibfnamefont {P.}~\bibnamefont {Lombardi}}, \bibinfo {author}
  {\bibfnamefont {I.}~\bibnamefont {Lomskaya}}, \bibinfo {author}
  {\bibfnamefont {L.}~\bibnamefont {Ludhova}}, \bibinfo {author} {\bibfnamefont
  {I.}~\bibnamefont {Machulin}}, \bibinfo {author} {\bibfnamefont
  {J.}~\bibnamefont {Martyn}}, \bibinfo {author} {\bibfnamefont
  {E.}~\bibnamefont {Meroni}}, \bibinfo {author} {\bibfnamefont
  {L.}~\bibnamefont {Miramonti}}, \bibinfo {author} {\bibfnamefont
  {M.}~\bibnamefont {Misiaszek}}, \bibinfo {author} {\bibfnamefont
  {V.}~\bibnamefont {Muratova}}, \bibinfo {author} {\bibfnamefont
  {R.}~\bibnamefont {Nugmanov}}, \bibinfo {author} {\bibfnamefont
  {L.}~\bibnamefont {Oberauer}}, \bibinfo {author} {\bibfnamefont
  {V.}~\bibnamefont {Orekhov}}, \bibinfo {author} {\bibfnamefont
  {F.}~\bibnamefont {Ortica}}, \bibinfo {author} {\bibfnamefont
  {M.}~\bibnamefont {Pallavicini}}, \bibinfo {author} {\bibfnamefont
  {L.}~\bibnamefont {Pelicci}}, \bibinfo {author} {\bibfnamefont
  {O.}~\bibnamefont {Penek}}, \bibinfo {author} {\bibfnamefont
  {L.}~\bibnamefont {Pietrofaccia}}, \bibinfo {author} {\bibfnamefont
  {N.}~\bibnamefont {Pilipenko}}, \bibinfo {author} {\bibfnamefont
  {A.}~\bibnamefont {Pocar}}, \bibinfo {author} {\bibfnamefont
  {G.}~\bibnamefont {Raikov}}, \bibinfo {author} {\bibfnamefont {M.~T.}\
  \bibnamefont {Ranalli}}, \bibinfo {author} {\bibfnamefont {G.}~\bibnamefont
  {Ranucci}}, \bibinfo {author} {\bibfnamefont {A.}~\bibnamefont {Razeto}},
  \bibinfo {author} {\bibfnamefont {A.}~\bibnamefont {Re}}, \bibinfo {author}
  {\bibfnamefont {N.}~\bibnamefont {Rossi}}, \bibinfo {author} {\bibfnamefont
  {S.}~\bibnamefont {Sch\"onert}}, \bibinfo {author} {\bibfnamefont
  {D.}~\bibnamefont {Semenov}}, \bibinfo {author} {\bibfnamefont
  {G.}~\bibnamefont {Settanta}}, \bibinfo {author} {\bibfnamefont
  {M.}~\bibnamefont {Skorokhvatov}}, \bibinfo {author} {\bibfnamefont
  {A.}~\bibnamefont {Singhal}}, \bibinfo {author} {\bibfnamefont
  {O.}~\bibnamefont {Smirnov}}, \bibinfo {author} {\bibfnamefont
  {A.}~\bibnamefont {Sotnikov}}, \bibinfo {author} {\bibfnamefont
  {R.}~\bibnamefont {Tartaglia}}, \bibinfo {author} {\bibfnamefont
  {G.}~\bibnamefont {Testera}}, \bibinfo {author} {\bibfnamefont
  {E.}~\bibnamefont {Unzhakov}}, \bibinfo {author} {\bibfnamefont {F.~L.}\
  \bibnamefont {Villante}}, \bibinfo {author} {\bibfnamefont {A.}~\bibnamefont
  {Vishneva}}, \bibinfo {author} {\bibfnamefont {R.~B.}\ \bibnamefont
  {Vogelaar}}, \bibinfo {author} {\bibfnamefont {F.}~\bibnamefont {von
  Feilitzsch}}, \bibinfo {author} {\bibfnamefont {M.}~\bibnamefont {Wojcik}},
  \bibinfo {author} {\bibfnamefont {M.}~\bibnamefont {Wurm}}, \bibinfo {author}
  {\bibfnamefont {S.}~\bibnamefont {Zavatarelli}}, \bibinfo {author}
  {\bibfnamefont {K.}~\bibnamefont {Zuber}}, \ and\ \bibinfo {author}
  {\bibfnamefont {G.}~\bibnamefont {Zuzel}} (\bibinfo {collaboration} {Borexino
  Collaboration}),\ }\href {\doibase 10.1103/PhysRevD.108.102005} {\bibfield
  {journal} {\bibinfo  {journal} {Phys. Rev. D}\ }\textbf {\bibinfo {volume}
  {108}},\ \bibinfo {pages} {102005} (\bibinfo {year} {2023})}\BibitemShut
  {NoStop}%
\end{thebibliography}
%

\end{document}